\documentclass[
final,3p,
12pt]{elsarticle}
\usepackage{lineno}

\usepackage[T1]{fontenc}
\usepackage{amsmath,amssymb,bm,mathtools}
\usepackage{graphicx}
\usepackage{xcolor}
\usepackage{siunitx}
\usepackage{booktabs}
\usepackage{url}
\usepackage{microtype}
\usepackage{algorithm}

\journal{Journal of Computational Physics}

\newcommand{\vect}[1]{\bm{#1}}
\newcommand{\E}{\vect{E}}
\newcommand{\B}{\vect{B}}
\newcommand{\J}{\vect{J}}
\newcommand{\Ji}{\vect{J}_{i}}
\newcommand{\Je}{\vect{J}_{e}}
\newcommand{\ue}{\vect{u}_{e}}
\newcommand{\rhoc}{\rho_c}

\newcommand{\Kop}{\mathcal{K}}
\newcommand{\I}{\mathsf{I}}
\newcommand{\Smat}{\mathsf{S}}
\newcommand{\Amat}{\mathsf{A}}
\newcommand{\Mmat}{\mathsf{M}}
\newcommand{\Hop}{\mathcal{H}}
\newcommand{\Dop}{\mathcal{D}}

\newcommand{\dE}{\delta\vect{E}}
\newcommand{\Wt}{\widehat{\vect{W}}}
\newcommand{\bhat}{\widehat{\vect{b}}}
\newcommand{\Cb}{\mathsf{C}_{\bhat}}
\newcommand{\ip}[2]{\langle #1,#2\rangle_h}

\begin{document}
\begin{frontmatter}

\title{A physics-based-preconditioned quasi-Newton solver for energy- and momentum-conserving implicit hybrid particle-in-cell simulations}

\author[helion]{Prabhat Kumar}
\author[helion]{Luis Chacón}
\author[helion]{William J. White}
\author[helion]{Lujain Almarhabi}
\author[helion]{Stephen E. Clark}
\address[helion]{Helion Energy, Inc., Everett, WA, USA}

\begin{abstract}
Explicit hybrid particle-in-cell (PIC) schemes, with kinetic ions and fluid electrons, accumulate numerical heating over long simulations and are unstable to finite-grid instabilities. Energy-conserving implicit integration removes heating and instabilities, but may be inefficient due to the need of nonlinear iteration. We present a fast quasi-Newton solver for an implicit hybrid-PIC formulation. The nonlinear residual employs the electric field and electron pressure as nonlinear unknowns, and nonlinearly eliminates particles. The formulation is strictly energy conserving and, on Cartesian meshes, also momentum conserving. The nonlinear iteration is efficiently performed with a quasi-Newton solver that replaces the true Jacobian with an approximate one without spoiling conservation properties upon convergence. The approximate Jacobian replaces the particle ion linear response with an analytical mass matrix that is local and requires minimal storage. This approximate Jacobian is efficiently inverted with a physics-based multigrid preconditioner, adapted from Hall MHD. We verify the scheme linearly with whistler, kinetic Alfv\'en, and cylindrical normal modes, and nonlinearly with the GEM reconnection challenge. On GEM, explicit integration needs a forty-times-smaller time step, and nine times the wall-clock time, to reach an energy conservation comparable to that of the implicit solver. We apply the method to the simulation of a three-dimensional field-reversed configuration formation by external coils.
\end{abstract}

\end{frontmatter}

\section{Introduction}
\label{sec:introduction}

Field-reversed configurations (FRCs) are compact, high-beta plasmas in which the ion gyroradius is comparable to the gradient scale lengths, so that ion kinetic effects influence stability and transport, while formation, merging, and compression unfold over many Alfv\'en transit times \cite{Steinhauer2011,Belova2003FRC}. Helion Energy's pulsed approach, in which FRCs are formed, merged, and magnetically compressed, is a demanding example \cite{Belova2026FRCMerging}. Hybrid particle-in-cell (PIC) methods, which treat the ions kinetically and the electrons as a fluid closed by a generalized Ohm's law, resolve the ion scales without the electron-plasma-frequency and light-speed constraints of fully kinetic PIC, and are widely used for such problems and for space and laboratory plasmas more broadly \cite{NielsonLewis1976,Winske2003,Groenewald2023,Belyaev2024}.

Despite its asymptotic reductions, hybrid PIC still faces a restrictive explicit time-step stability limit. The Hall term in Ohm's law supports whistler waves whose frequency grows as $k^2$ at short wavelengths, requiring $\Delta t\propto\Delta x^2$ \cite{Huba1995,Chacon2003}. With explicit integration, halving the mesh spacing reduces the stable time step fourfold, and the sub-cycling typically used to relieve the constraint accumulates error over long simulations. Explicit hybrid schemes typically conserve neither energy nor momentum and, under certain conditions, are subject to the finite-grid instability~\cite{Rambo1995}. Its growth rate is not arrested by mesh refinement, resulting in secular numerical heating.
Fully implicit, energy-conserving integration removes stability constraints and secular heating \cite{Stanier2019,Kumar2026ThetaImplicitHybrid}, and lets the time step follow the ion dynamics. Its wall-clock-time performance then rests on the cost of the nonlinear solver. In a Jacobian-free Newton--Krylov (JFNK) method \cite{KnollKeyes2004}, every Krylov iteration re-evaluates the nonlinear residual and therefore re-advances all particles, so that its cost is approximately proportional to the number of Krylov iterations, which grows with the condition number of the Jacobian unless the linear system is preconditioned. Even with efficient preconditioning, JFNK still requires a particle push per Krylov iteration, which can remain too expensive in practice and motivates using a quasi--Newton approach that employs a suitable approximation to the Jacobian that avoids the particles altogether, e.g. by using an analytical mass-matrix approximation of the particle response \cite{Angus2022,Angus2023}.

Conservative implicit hybrid PIC has a firm theoretical foundation. Stanier et al.\ combined the implicit midpoint rule, nonlinear elimination of the particles, and JFNK into a fully implicit hybrid algorithm that conserves energy and, on Cartesian meshes, momentum \cite{Stanier2019}. The Cartesian implementation was later extended to curvilinear meshes \cite{Stanier2022}. 
A $\theta$-implicit hybrid-PIC advance in WarpX \cite{Kumar2026ThetaImplicitHybrid} employed the electric field of Ohm's law as the sole unknown, adapted the particle-based mass matrices of energy-conserving semi-implicit PIC \cite{Lapenta2017}, extended to the fully implicit setting by Angus et al.\ \cite{Angus2022,Angus2023}, to approximate the Ohm's-law Jacobian, and isolated the unpreconditioned Krylov solve as the dominant cost at large time steps.

For hyperbolic PDEs, physics-based preconditioning is the established route to manageable Krylov counts that grow slowly with the time step \cite{KnollKeyes2004}. In implicit MHD, approximate block factorizations reduce the coupled system to Schur complements that multigrid inverts efficiently \cite{ChaconKnoll2002,Chacon2008}. In implicit PIC, moment-based preconditioners play the same role for the kinetically enslaved JFNK solve, first in the electrostatic model \cite{Chen2014FluidPC} and then in the Darwin model \cite{Chen2015Darwin}. Most directly relevant here, the fully implicit Hall-MHD solver in \cite{Chacon2025HMHD} isolates the stiff electron block, poses it for the vector potential, and relaxes it with a coupled smoother that inverts a local approximation at each node while the multigrid residual retains the exact discrete curl--curl operator.

Our focus is to extend this Hall-MHD preconditioner to kinetic-ion hybrid PIC. 
%
As in \cite{Chacon2025HMHD}, its field block couples the electric-field update to an auxiliary curl--curl variable, and it retains the Hall rotation, resistive and hyper-resistive damping, and electron convection. Beyond the reference, we adapt it to the hybrid PIC context by employing a local approximation of the ion linear response via analytical mass matrices, discussed below. As in \cite{Chacon2025HMHD}, relaxation is by a block-Jacobi smoother whose nodal block has a closed-form inverse, inside a geometric multigrid $V$-cycle. A block-triangular pressure solve provides the electron-pressure response to transport and conduction and couples it to the field through the pressure gradient. The field block is solved approximately using multigrid-preconditioned inner GMRES within an outer flexible GMRES (FGMRES) solver.

While the preconditioner reduces the number of Krylov iterations, in JFNK the cost of each iteration is still set by the ion response in the Jacobian action. Particle-based mass matrices \cite{Lapenta2017} remove the particle advance from the Krylov iterations, but they are deposited by a particle loop at every Newton iteration and stored as a $3\times3$ block for each pair of nodes within the shape-function support, 243 values per node in three dimensions with linear shapes. To eliminate this cost, we follow \cite{Chen2015Darwin} and employ an analytical mass matrix (AMM) approximation to the ion response in the Jacobian matrix. 
The resulting approximate Jacobian action employs no particle loop and only needs to store four values per node, and is thus extremely memory and cost friendly. 

Since the particle response in the Newton residual is preserved, neither the converged solution nor its conservation of energy and momentum are affected by the Jacobian approximations. Total energy is conserved at the implicit midpoint to nonlinear and particle-solve accuracy. On periodic Cartesian meshes, total momentum is conserved as well (Section~\ref{subsec:gem}). The method is implemented in the open-source code WarpX \cite{Vay2018WarpX} and builds on AMReX's infrastructure for GPU acceleration, performance portability across CPU and GPU architectures, and deployment on exascale systems \cite{Zhang2019AMReX}.

The implementation is verified with parallel whistler and oblique kinetic Alfv\'en waves, which are reproduced to better than $0.1\%$ in frequency against the cold-ion linear solution. It is also verified against the normal-mode spectrum of a plasma-filled conducting cylinder, where the computed spectrum follows the cold-plasma fast and slow branches at a dimensionless grid-scale whistler stiffness exceeding ten. On the GEM reconnection problem, the reconnected flux and the transfer of magnetic to ion energy reproduce the fully implicit reference of Stanier et al.\ \cite{Stanier2019}. The total-energy error stays below the Newton tolerance in every run; with that tolerance tightened to $10^{-13}$, total energy is conserved to roundoff and total momentum to better than $10^{-9}$ in Alfv\'enic units. At $\Delta t\Omega_{ci}=0.05$, the preconditioner reduces the outer Krylov iterations per Newton iteration from 357 to about 6, with only a weak dependence on the time step. Together with the analytical mass matrices, it reduces the time per step over the first 25 steps 186-fold relative to unpreconditioned JFNK, and twofold relative to using particle-based mass matrices with the same preconditioner. Over the full reconnection interval, an explicit hybrid-PIC calculation at the implicit time step, with field substeps at its whistler stability limit, is 2.8 times faster but has a total-energy error more than four orders of magnitude larger; reaching a comparable energy error requires a forty-times-smaller time step and nine times the implicit wall-clock time. Finally, we apply the method to the three-dimensional formation of an FRC by external coils and study the formation dynamics.

The rest of the paper is organized as follows. Section~\ref{sec:method} states the implicit formulation and its conservation properties, develops the preconditioner, and introduces the analytical mass matrix approximation to the ion linear response. Section~\ref{sec:verification} verifies the solver on linear wave propagation, cylindrical normal modes, and the GEM reconnection problem, and compares its cost with standard JFNK and with explicit integration. Section~\ref{sec:frc} applies it to three-dimensional FRC formation, and Section~\ref{sec:conclusions} summarizes the findings.

\section{Hybrid-PIC formulation and accelerated Newton--Krylov solver}
\label{sec:method}

We first summarize the discrete hybrid-PIC equations and their conservation properties, then describe the quasi-Newton-Krylov solve and the stiffness that motivates preconditioning. The $\theta$-implicit advance follows Ref.~\cite{Kumar2026ThetaImplicitHybrid}. Sections~\ref{subsec:pcapprox} and~\ref{subsec:fieldblock} develop the physics-based preconditioner, which reduces the number of Krylov iterations, and Section~\ref{subsec:fluid} removes the particle work from each Krylov iteration with the analytical mass matrix ion response. Additional block algebra is provided in \ref{app:pcderivation}. Unless stated otherwise, the equations use a collocated Cartesian mesh, as in Refs.~\cite{Stanier2019,Stanier2022}, in one, two, or three spatial dimensions, with three field components; the collocated cylindrical extension, restricted to the azimuthally symmetric mode, is noted where it differs.

\subsection{Hybrid-PIC model and $\theta$-implicit advance}
\label{subsec:model}

In the nonrelativistic limit, the particle ion equations of motion are
\begin{equation}
 \frac{d\vect{x}_p}{dt}=\vect{v}_p,
 \qquad
 \frac{d\vect{v}_p}{dt}=\frac{q_p}{m_p}
 \left[\E^*(\vect{x}_p)+\vect{v}_p\times\B(\vect{x}_p)\right],
 \label{eq:particle}
\end{equation}
where $\E^*=\E-\vect{D}$ is the ion push field and $\vect{D}$ contains the dissipative terms assigned to the electron fluid (defined below). Quasineutrality and Ampère's law without displacement current give
\begin{equation}
 \rhoc=e n_e>0,\qquad
 \J=\frac{1}{\mu_0}\nabla_h\times\B,\qquad
 \Je=\J-\Ji,
 \label{eq:ampere}
\end{equation}
with $\ue=-\Je/\rhoc$. Here $e$ is the elementary charge, $\rhoc$ and $\Ji$ are the deposited ion charge and current densities, and $\nabla_h$ denotes the appropriate discrete derivative. For inertialess electrons, the generalized Ohm's law is
\begin{equation}
 \E=\frac{\Je\times\B-\nabla_h P_e}{\rhoc}+\vect{D},
 \qquad
 \vect{D}=\eta\J-\eta_h\nabla_h^2\J.
 \label{eq:ohm}
\end{equation}
The resistivity $\eta(\rhoc,|\J|,t)$ and hyper-resistivity $\eta_h(\rhoc,|\B|)$ are scalar functions evaluated at each node. On the Cartesian mesh, the hyper-resistive term uses a compact componentwise Laplacian; on the cylindrical $r-z$ mesh, it uses the components of the azimuthally symmetric vector Laplacian, with a control-volume form on the axis. 

The magnetic field is updated from Faraday's law:
\begin{align}
 \B^{n+1}&=\B^n-\Delta t\,\nabla_h\times\E^{n+\theta},
 \label{eq:faraday}\\
 \B^{n+\theta}&=(1-\theta)\B^n+\theta\B^{n+1},
 \qquad \tfrac12\leq\theta\leq1.
 \nonumber
\end{align}
The particles are advanced by the implicit midpoint rule with the fields at $n+\theta$,
\begin{equation}
 \vect{x}_p^{n+1}=\vect{x}_p^n+\Delta t\,\bar{\vect{v}}_p,
 \qquad
 \vect{v}_p^{n+1}=\vect{v}_p^n+\Delta t\,\frac{q_p}{m_p}\left(\E^*_p+\bar{\vect{v}}_p\times\B_p\right),
 \label{eq:push}
\end{equation}
where $\bar{\vect{x}}_p=(\vect{x}_p^n+\vect{x}_p^{n+1})/2$ and $\bar{\vect{v}}_p=(\vect{v}_p^n+\vect{v}_p^{n+1})/2$. Fields are gathered to the particles, and moments are deposited to the mesh, at the midpoint positions with the same shape function $S(\vect{x})=\prod_dS_s(x_d/\Delta x_d)$, where $S_s$ is the centered B-spline of degree $s$ ($s=1$ for linear shapes):
\begin{gather}
 \E^*_p=\sum_g S(\bar{\vect{x}}_p-\vect{x}_g)\,(\mathsf{F}\E^*)_g,
 \qquad
 \B_p=\sum_g S(\bar{\vect{x}}_p-\vect{x}_g)\,(\mathsf{F}\B)_g,
 \label{eq:gather}\\
 (\rho_{\rm dep})_g=\frac{1}{\Delta V_g}\sum_pq_p\,S(\bar{\vect{x}}_p-\vect{x}_g),
 \qquad
 (\J_{\rm dep})_g=\frac{1}{\Delta V_g}\sum_pq_p\bar{\vect{v}}_p\,S(\bar{\vect{x}}_p-\vect{x}_g),
 \nonumber\\
 \rhoc=\mathsf{F}\rho_{\rm dep},\qquad \Ji=\mathsf{F}\J_{\rm dep},
 \label{eq:deposit}
\end{gather}
where $\Delta V_g$ is the control volume of node $g$. The operator $\mathsf{F}$ is the identity or, on periodic meshes, a symmetric binomial filter applied identically to the gathered fields and the deposited moments, so that the gather is the transpose of the deposition. 

Prescribed external fields, such as those of formation and compression coils, enter through the exact vector-potential splitting of Ref.~\cite{Kumar2026ThetaImplicitHybrid}: $\B=\B_{\rm int}+\B_{\rm ext}$ and $\E=\E_{\rm int}+\E_{\rm ext}$. For $N$ prescribed coils, the external vector potential is
\begin{equation*}\label{eq:vector_potential}
 \vect{A}_{\rm ext}(\vect{x},t)
 =\sum_{c=1}^{N}\vect{A}_c(\vect{x})f_c(t),
\end{equation*}
where $\vect{A}_c$ describes each coil's spatial profile and $f_c$ its driving waveform. The external fields are $\B_{\rm ext}=\nabla\times\vect{A}_{\rm ext}$ and $\E_{\rm ext}=-\partial_t\vect{A}_{\rm ext}$. The prescribed external electric and magnetic fields satisfy Faraday's law identically, so Eq.~\eqref{eq:faraday} advances $\B_{\rm int}$ without splitting error. The plasma current is $\nabla_h\times\B_{\rm int}/\mu_0$, the coil current being excluded from Eq.~\eqref{eq:ampere}, while the Hall term and the particle push use the total electromagnetic fields. 

The electron pressure, with $\gamma>1$, satisfies the discrete work equation
\begin{equation}
 \begin{aligned}
 \frac{P_e^{n+\theta}-P_e^n}{\theta\Delta t}
 ={}&-\gamma\nabla_h\cdot(\ue P_e^{n+\theta})\\
 &+(\gamma-1)\left[\Je\cdot\E^*+Q_e
 +\nabla_h\cdot\left(\kappa_e\nabla_h\frac{P_e^{n+\theta}}{n_e}\right)\right].
 \end{aligned}
 \label{eq:pe}
\end{equation}
The field and fluid quantities on the right are evaluated at the current $n+\theta$ state, with $\Ji$ deposited from the midpoint particle advance as described earlier. The temperature $P_e/n_e$ is in energy units. At an Ohm's-law solution, $\Je\cdot\E^*=\ue\cdot\nabla_hP_e$; writing the work as $\Je\cdot\E^*$ in Eq.~\eqref{eq:pe} makes it consistent with the discrete particle and field work. The pressure at $n+1$ follows by $\theta$ extrapolation. Transport and conduction use conservative face fluxes, and $Q_e$ is the dissipative heating, whose discrete form is given in Section~\ref{subsec:energy}.

The pressure flux uses a second-order upwind (MUSCL) face reconstruction with the van Albada slope limiter~\cite{VanAlbada1982}. In this flux, $\ue$ is replaced by an advection-limited velocity $\vect{u}_c$, bounded smoothly at a fraction of the mesh Courant speed $\Delta x_{\min}/(\theta\Delta t)$; the bound acts inside the conservative flux, so the transport term still telescopes when volume integrated. A density floor regularizes $1/\rhoc$ and a positivity floor acts on the pressure; where active, the floors change the discrete equations and the energy balance of Section~\ref{subsec:energy}.

The electric field and electron pressure are the Newton unknowns, giving the coupled mesh system
\begin{equation}
 \vect{U}=\begin{bmatrix}\E^{n+\theta}\\P_e^{n+\theta}\end{bmatrix},
 \qquad
 \vect{F}(\vect{U})=
 \begin{bmatrix}\vect{F}_E(\E,P_e)\\F_P(\E,P_e)\end{bmatrix}=\vect{0},
 \label{eq:nonlinear}
\end{equation}
where $\vect{F}_E=\E-\E_{\rm Ohm}[\B(\E),\Ji(\E),\rhoc(\E),P_e]$. We normalize $F_P$ by multiplying Eq.~\eqref{eq:pe}, with all terms moved to the left, by $\theta\Delta t$, so that its time-derivative contribution is $P_e^{n+\theta}-P_e^n$. The particle unknowns and the magnetic field are nonlinearly eliminated~\cite{Chen2011,Chen2015Darwin,Stanier2019,Angus2023,Kumar2026ThetaImplicitHybrid}: for each trial mesh state, their implicit equations, Eq.~\eqref{eq:push}, are solved by a per-particle Picard iteration before the moments are deposited. Thus, the Newton system contains only mesh unknowns, yet its solution retains the self-consistent particle response.

\subsection{Discrete energy and momentum conservation}
\label{subsec:energy}

The appendix of Ref.~\cite{Kumar2026ThetaImplicitHybrid} derives the discrete particle and magnetic work identities, including the splitting for prescribed fields. We retain here the resulting balance and the electron-heating condition needed to close it. Using the mesh inner product
\[
 \ip{\vect{a}}{\vect{b}}=\sum_g\vect{a}_g\cdot\vect{b}_g\Delta V_g,
\]
with each periodic node counted once, the total energy is
\begin{equation}
 W_{\rm tot}=W_i+W_B+W_e,\qquad
 W_B=\frac{\ip{\B_{\rm int}}{\B_{\rm int}}}{2\mu_0},\qquad
 W_e=\sum_g\frac{(P_e)_g}{\gamma-1}\Delta V_g.
 \label{eq:energydefs}
\end{equation}
Here $W_i$ includes the macro-particle weights and $\B_{\rm int}$ is the internal magnetic field. The transposed gather and deposition of Eqs.~\eqref{eq:gather} and~\eqref{eq:deposit}, summation by parts for the discrete curls, and conservative pressure and heat fluxes give the ion and electron work contributions $\ip{\Ji}{\E^*}$ and $\ip{\Je}{\E^*}+\sum_g(Q_e)_g\Delta V_g$. Their sum cancels the magnetic work $-\ip{\J}{\E}$ when
\begin{equation}
 \sum_g(Q_e)_g\Delta V_g=\ip{\vect{D}}{\J}.
 \label{eq:heatingbalance}
\end{equation}
The positive-definite heating source used here is
\begin{equation}
 (Q_e)_g=\eta|\J_g|^2
 +\eta_h\sum_{c,d}\bigl\langle(\partial_dJ_c)^2\bigr\rangle_g,
 \label{eq:Qpos}
\end{equation}
where the brackets average squared one-sided differences over the two faces adjacent to node $g$ in direction $d$. The resistive part satisfies Eq.~\eqref{eq:heatingbalance} node by node for any $\eta$. For constant $\eta_h$, the hyper-resistive part is the summation-by-parts partner of the compact Laplacian in Eq.~\eqref{eq:ohm}, and Eq.~\eqref{eq:heatingbalance} holds exactly; spatial variation of $\eta_h$ makes this equality approximate.

For a closed system, the resulting energy law is
\begin{equation}
 W_{\rm tot}^{n+1}-W_{\rm tot}^n
 =-\frac{\theta-1/2}{\mu_0}
 \left\|\B^{n+1}-\B^n\right\|_h^2.
 \label{eq:energyfinal}
\end{equation}
Equation~\eqref{eq:energyfinal} assumes Eq.~\eqref{eq:heatingbalance} and that pressure floors, vacuum modifications, and other energy-changing corrections are inactive. At $\theta=1/2$ the total energy is conserved to nonlinear and particle-solve accuracy; $\theta>1/2$ adds only the displayed magnetic dissipation. Boundary energy fluxes and prescribed-field input must be included for an open or driven system, as detailed in Ref.~\cite{Kumar2026ThetaImplicitHybrid}.

The electrons are massless, so the total momentum is the ion momentum $\vect{P}_i=\sum_pm_p\vect{v}_p$, with the macro-particle weights included in $m_p$ and $q_p$. The push, Eq.~\eqref{eq:push}, changes $\vect{P}_i$ through the forces on the midpoint orbits. Since the gather is the transpose of the deposition for both $\E^*$ and $\B$, the sum of these forces over all ions equals a sum over mesh nodes.  At an Ohm's-law solution, $\rhoc\E^*=\Je\times\B-\nabla_hP_e$, the dissipative terms cancel between ions and electrons, and
\begin{equation}
 \vect{P}_i^{n+1}-\vect{P}_i^n
 =\Delta t\sum_p q_p\bigl[\E^*_p+\bar{\vect{v}}_p\times\B_p\bigr]
 =\Delta t\sum_g\bigl(\J\times\B-\nabla_hP_e\bigr)_g\Delta V_g,
 \label{eq:momentum}
\end{equation}
with the fields at $n+\theta$. On a periodic Cartesian mesh with centered differences, $\sum_g(\nabla_hP_e)_g\Delta V_g=\vect{0}$ and $\mu_0\sum_g(\J\times\B)_g\Delta V_g=-\sum_g\B_g(\nabla_h\cdot\B)_g\Delta V_g$, which vanishes when the initial field satisfies $\nabla_h\cdot\B=0$, since Eq.~\eqref{eq:faraday} preserves $\nabla_h\cdot\B$. Total momentum is then conserved for any $\theta$, to nonlinear and particle-solve accuracy, provided the density floor is inactive. With conducting walls or prescribed coil fields, the plasma exchanges momentum with the boundaries and the coils, and only the corresponding balance holds. Both balances concern the converged discrete equations, independently of how the Newton updates are computed, and Section~\ref{subsec:gem} measures them on the GEM problem.

\subsection{Quasi-Newton-Krylov solve}
\label{subsec:newton}

Equation~\eqref{eq:nonlinear} is solved by an inexact Newton iteration in which a Krylov method computes each update from an approximate Jacobian $\widetilde{\mathsf{J}}_k$:
\begin{equation}
 \widetilde{\mathsf{J}}_k\delta\vect{U}_k=-\vect{F}_k,
 \qquad \vect{U}_{k+1}=\vect{U}_k+\lambda_k\delta\vect{U}_k,
 \label{eq:newton}
\end{equation}
where $\vect{F}_k=\vect{F}(\vect{U}_k)$. The iteration stops when $\|\vect{F}_k\|\leq\epsilon_t=\epsilon_r\|\vect{F}_0\|+\epsilon_a$, with relative and absolute tolerances $\epsilon_r$ and $\epsilon_a$, and convergence is always assessed on the full nonlinear residual, including the implicit particle advance. The step length $\lambda_k$ is the first of $1,1/2,\dots,2^{-6}$ that reduces $\|\vect{F}\|$; if none does, the update is discarded and recomputed with the linear tolerance $\zeta_{\min}$ defined below, and a second failure ends the step. Each linear solve is inexact, with a fixed relative tolerance or the adaptive forcing of Eisenstat and Walker~\cite{EisenstatWalker1996} with the safeguards of Ref.~\cite{Chacon2025HMHD}:
\begin{equation}
 \begin{aligned}
 \zeta_k&=\max\bigl\{\zeta_{\min},\,\min\bigl[\zeta_{\max},\,
 \max\bigl(\zeta_k^{A},\,\gamma_f\zeta_{k-1}^{\,\xi},\,\gamma_f\epsilon_t/\|\vect{F}_k\|\bigr)\bigr]\bigr\},\\
 \zeta_k^{A}&=\gamma_f\bigl(\|\vect{F}_k\|/\|\vect{F}_{k-1}\|\bigr)^{\xi},
 \end{aligned}
 \label{eq:forcing}
\end{equation}
with $\zeta_0=\zeta_{\max}$, $\gamma_f=0.9$, $\xi=1.5$, $\zeta_{\max}=0.8$, and $\zeta_{\min}=10^{-4}$. The Krylov solver is right-preconditioned: it solves $(\widetilde{\mathsf{J}}_k\mathsf{P}^{-1})(\mathsf{P}\,\delta\vect{U}_k)=-\vect{F}_k$ with a preconditioner $\mathsf{P}\approx\widetilde{\mathsf{J}}_k$ that is cheap to invert.

The exact Jacobian $\mathsf{J}_k$ of $\vect{F}$ contains the response of every particle orbit to the field update. In a Jacobian-free Newton--Krylov (JFNK) method~\cite{KnollKeyes2004}, its action is the finite difference $\mathsf{J}_k\vect{v}\approx[\vect{F}(\vect{U}_k+\varepsilon\vect{v})-\vect{F}(\vect{U}_k)]/\varepsilon$, so every Krylov iteration re-advances all particles, and a Newton iteration costs the number of Krylov iterations times a particle advance; Section~\ref{sec:performance} uses JFNK as the reference. The approximate Jacobian $\widetilde{\mathsf{J}}_k$ instead freezes the particle orbits at $\vect{U}_k$: the ion current responds to the push-field update through a mass matrix (Section~\ref{subsec:fluid}), and the action of $\widetilde{\mathsf{J}}_k$ is evaluated matrix-free, as a directional difference of the residual in which the particle advance is replaced by this linear response. Equation~\eqref{eq:newton} is therefore a quasi-Newton method. The terms omitted by freezing the orbits remain in $\vect{F}$, which advances the particles at every Newton iterate, so the approximation affects the rate of convergence but neither the converged solution nor its conservation properties (Section~\ref{subsec:energy}).

The number of Krylov iterations is set by the stiff field response: in particular, the implicit Hall response stiffness grows as $\Delta t/\Delta x^2$, increasing the Krylov work as the step is enlarged or the mesh refined. Sections~\ref{subsec:pcapprox} and~\ref{subsec:fieldblock} derive a physics-based multigrid preconditioner that reduces the number of Krylov iterations by approximately inverting the stiff field response: the Hall, dissipative, and electron-convection terms, the local ion current response, and electron-pressure transport. Section~\ref{subsec:fluid} computes the mass matrix analytically on the mesh, so that a Krylov iteration requires no particle operations.

\subsection{Physics-based preconditioner}
\label{subsec:pcapprox}
The Newton system, Eq.~\eqref{eq:newton}, has the block structure
\begin{equation}
 \begin{bmatrix}J_{EE}&G_P\\C_{PE}&J_{PP}\end{bmatrix}
 \begin{bmatrix}\dE\\\delta P_e\end{bmatrix}
 =\begin{bmatrix}\vect{b}_E\\b_P\end{bmatrix},
 \qquad G_P\delta P_e=\frac{\nabla_h\delta P_e}{\rhoc},
 \label{eq:coupledblocks}
\end{equation}
where $G_P$ is the pressure-gradient coupling of Ohm's law and $C_{PE}$ contains the response of the electron velocity, work, heating, and density in Eq.~\eqref{eq:pe} to the electric-field update. As in the Hall-MHD solver of Ref.~\cite{Chacon2025HMHD} (Sec.~3.2), we drop $C_{PE}$ and approximate the two diagonal blocks, which gives the block-triangular preconditioner
\begin{equation}
 \mathsf{P}=\begin{bmatrix}\mathcal{J}_E&G_P\\0&L_P\end{bmatrix},
 \qquad
 \begin{aligned}
 L_P\delta P_e&=b_P,\\
 \mathcal{J}_E\dE&=\vect{b}_E-G_P\delta P_e.
 \end{aligned}
 \label{eq:triangularpc}
\end{equation}
Each application of $\mathsf{P}^{-1}$ solves the pressure block first and then the field block, whose right-hand side contains the gradient of the pressure update. The field block $\mathcal{J}_E$ approximates $J_{EE}$, and $L_P$ approximates $J_{PP}$. Both are built from the state at the current Newton iterate, denoted by the subscript ``0'': the density $\rho_{c,0}$, the total magnetic field $\B_0$, the electron velocity $\vect{u}_{e,0}$ of Ohm's law, and $\eta$ and $\eta_h$ evaluated with $\rho_{c,0}$, $|\J_0|$, and $|\B_0|$. These coefficients vary from cell to cell, but their updates are omitted.

For the field block, the pressure is held fixed. The magnetic field responds through the implicit Faraday update,
\begin{equation}
 \delta\B^{n+\theta}=-\theta\Delta t\,\nabla_h\times\dE,
 \label{eq:dB}
\end{equation}
and the linearization of Ohm's law has four contributions: the updates of the plasma current, the magnetic field, the dissipative field, and the deposited ion current. We define
\begin{equation}
 \begin{gathered}
 \Kop=\nabla_h\times\nabla_h\times,\qquad
 \vect{w}=\theta\Delta t\,\vect{u}_{e,0},\qquad
 \bhat=\frac{\B_0}{|\B_0|},\qquad
 \Cb\vect{v}=\vect{v}\times\bhat,\\
 \Hop=c_w\Cb,\qquad
 c_w=\frac{\theta\Delta t\,|\B_0|}{\mu_0\rho_{c,0}},\qquad
 \Dop\vect{v}=a_\eta\vect{v}-a_h\nabla_h^2\vect{v},\\
 a_\eta=\frac{\theta\Delta t\,\eta}{\mu_0},\qquad
 a_h=\frac{\theta\Delta t\,\eta_h}{\mu_0},
 \end{gathered}
 \label{eq:Kandw}
\end{equation}
with $\bhat=\vect{0}$ at magnetic nulls. Through Ampère's law, $\delta\J=-(\theta\Delta t/\mu_0)\Kop\dE$, so the plasma-current part of the Hall term responds as $\delta\J\times\B_0/\rho_{c,0}=-\Hop\Kop\dE$. The frozen electron current acting on $\delta\B$ gives the convection term $\J_{e,0}\times\delta\B/\rho_{c,0}=\vect{w}\times(\nabla_h\times\dE)$, and the dissipative field responds as $\delta\vect{D}=-\Dop\Kop\dE$.

The ions are pushed with $\E^*=\E-\vect{D}$, so the deposited ion current responds to the push-field update $\delta\E^*=\dE+\Dop\Kop\dE$. With the particle orbits frozen at the current Newton iterate, this response is linear, $\delta\Ji=\mathsf{F}\Mmat\mathsf{F}\,\delta\E^*$, where the $3\times3$ blocks $\Mmat_{gg'}$ of the mass matrix, given in Section~\ref{subsec:fluid}, couple nodes within the support of the particle shape function. The preconditioner omits the filter and keeps only the component diagonal of the same-node block,
\begin{equation}
 \boldsymbol{\Sigma}_g=\operatorname{diag}(\Mmat_{gg}),
 \qquad
 \Smat\vect{v}=\frac{(\boldsymbol{\Sigma}\vect{v})\times\B_0}{\rho_{c,0}},
 \qquad
 \Amat=\I+\Smat.
 \label{eq:Sdef}
\end{equation}
$\boldsymbol{\Sigma}$ is a local ion conductivity: the current driven at a node, over the half step, by the push field at the same node. The ion part of the Hall term then responds as $-\delta\Ji\times\B_0/\rho_{c,0}=-\Smat(\dE+\Dop\Kop\dE)$. Although $\boldsymbol{\Sigma}$ is diagonal, the cross product makes $\Smat$, and hence $\Amat$, a coupled $3\times3$ matrix at each node.

Collecting the four contributions in $\delta\vect{F}_E=\dE-\delta\E_{\rm Ohm}$ gives the field block (\ref{app:fieldchain})
\begin{equation}
 \mathcal{J}_E\dE=\Amat\dE+(\Hop+\Dop+\Smat\Dop)\Kop\dE-\vect{w}\times(\nabla_h\times\dE).
 \label{eq:pc-field}
\end{equation}
Equation~\eqref{eq:pc-field} has the structure of the electron block of the Hall-MHD preconditioner of Ref.~\cite{Chacon2025HMHD} (Sec.~4.1): $\Hop\Kop$ is the stiff whistler response, $\Dop\Kop$ the resistive and hyper-resistive damping, and the last term the frozen electron convection. Compared with Hall MHD, the kinetic ions add two terms: $\Smat\dE$, which enters through $\Amat=\I+\Smat$, and $\Smat\Dop\Kop\dE$. The first is the ion-current response to the electric field. The second is the ion-current response to the dissipative part $\vect{D}$ of the push field $\E^*=\E-\vect{D}$.

For the pressure block, the electric field and density are held fixed in Eq.~\eqref{eq:pe}. The work and heating terms then have no direct pressure dependence, and freezing the advecting velocity and conductivity leaves the transport and conduction terms, which we approximate by
\begin{equation}
 L_P=I+\theta\Delta t
 \left[\gamma\vect{u}_{c,0}\cdot\nabla_h^{\rm up}
 -(\gamma-1)\frac{\kappa_0}{n_{e,0}}\nabla_h^2\right],
 \label{eq:pressureop}
\end{equation}
discretized by first-order upwind advection and the compact Laplacian, respectively (\ref{app:pressure}). The upwind stencil is a lower-order approximation of the limited second-order flux of the residual (Section~\ref{subsec:model}), as is common in physics-based preconditioning~\cite{KnollKeyes2004}. Here $\vect{u}_c$ is the advection-limited velocity of the pressure flux (Section~\ref{subsec:model}), and $\kappa_0$ is a representative scalar conductivity evaluated at the model's reference density and temperature. The omitted terms, the compressibility $\gamma(\nabla\cdot\vect{u}_c)\delta P_e$ and the derivatives of density and conductivity inside the heat flux, contain no derivatives of $\delta P_e$ and do not contribute to the stiffness of the pressure block. The pressure update is initialized with $\delta P_e=D_P^{-1}b_P$, where
\begin{equation}
 D_P=1+\theta\Delta t
 \left[\gamma\sum_d\frac{|u_{c,d}|}{\Delta x_d}
 +(\gamma-1)\frac{\kappa_0}{n_{e,0}}\sum_d\frac{2}{\Delta x_d^2}\right],
 \label{eq:Dp}
\end{equation}
is the diagonal of $L_P$. Jacobi sweeps, $\delta P_e\leftarrow\delta P_e+D_P^{-1}(b_P-L_P\delta P_e)$, then add the neighboring transport and conduction contributions. In stiffer contexts, the pressure block can instead be solved with multigrid, as in Ref.~\cite{Chacon2025HMHD}.

The preconditioner variables take the boundary conditions of the fields they update. The electric-field update $\dE$ satisfies those of $\E$: at a perfectly conducting face its tangential components vanish, so their rows are identity rows, and its ghost values are mirrored with the parities of $\E$. The magnetic update $\delta\B=-\theta\Delta t\nabla_h\times\dE$ and the current update $\delta\J\propto\Kop\dE$, and hence the auxiliary variable of Section~\ref{subsec:fieldblock}, take the parities of $\B$ and $\J$, which fix the ghost values of the curls in Eq.~\eqref{eq:pc-field}. At a symmetry (perfect magnetic conductor) face, the parities of the electric and magnetic fields are interchanged. On the collocated mesh, the residual takes the normal component of $\E$ at a conducting-wall node from the adjacent interior node rather than from Ohm's law, and that row is an identity row in the preconditioner as well. Nodes covered by an embedded boundary, where $\E$ is not updated, are identity rows. In cylindrical geometry, the axis is a symmetry boundary with the parities of the azimuthally symmetric mode, and the outer wall is a perfect conductor. The pressure update is periodic, or even across the axis; at other boundaries, the preconditioner extends its variables by zero.

\subsection{Multigrid solution of the field block}
\label{subsec:fieldblock}
With hyper-resistivity, Eq.~\eqref{eq:pc-field} contains the fourth-order term $\Dop\Kop$. As in Ref.~\cite{Chacon2025HMHD} (Sec.~4.1), we introduce the auxiliary variable
\[
 \vect{W}=\Kop\dE=-\frac{\mu_0}{\theta\Delta t}\,\delta\J,
\]
which is proportional to the plasma-current update, and solve for the electric-field update and $\vect{W}$ together:
\begin{equation}
 \begin{aligned}
 \Amat\dE+(\Hop+\Dop+\Smat\Dop)\vect{W}-\vect{w}\times(\nabla_h\times\dE)&=\vect{b}_E-G_P\delta P_e,\\
 \vect{W}-\Kop\dE&=\vect{0}.
 \end{aligned}
 \label{eq:unscaledfield}
\end{equation}
Every row now contains at most second spatial derivatives, so the three components of $\dE$ and the three of $\vect{W}$ at a node can be relaxed together with a local block that captures the Hall rotation. The auxiliary variable belongs only to the preconditioner and does not enlarge the Newton unknown vector.

The local block replaces derivative couplings by the scales
\begin{equation}
 \ell_1=\sum_d\frac{2}{\Delta x_d},
 \qquad
 \ell_2=\sum_d\frac{2}{\Delta x_d^2}
 \label{eq:l2cart}
\end{equation}
of the first- and second-derivative operators. On the Cartesian mesh, $\ell_2$ is the diagonal of the negative compact Laplacian, whereas the centered curl composition $\Kop$ has largest eigenvalue $\ell_2/2$. On the cylindrical mesh, the residual composes compact curl differences, and the block uses the corresponding scales of the cylindrical operators, including their on-axis forms.

Two normalizations precede the relaxation. First, the field row is multiplied by $\Amat^{-1}$, which places the identity on its diagonal. $\Amat$ has a closed-form inverse at each node; for an isotropic $\boldsymbol{\Sigma}$, with $\overline{m}=|\B_0|\operatorname{tr}\boldsymbol{\Sigma}/(3\rho_{c,0})$,
\begin{equation}
 \Amat^{-1}\vect{v}=\frac{\vect{v}-\overline{m}\,\vect{v}\times\bhat+\overline{m}^2(\vect{v}\cdot\bhat)\bhat}{1+\overline{m}^2},
 \label{eq:Ainv}
\end{equation}
and the anisotropic case uses the adjugate of the $3\times3$ matrix. Second, the auxiliary variable is stored scaled, $\Wt=s_W\vect{W}$, with the constant
\begin{equation}
 s_W=\max\Bigl(\max_g c_w,\ \ell_2^{-1}\Bigr).
 \label{eq:wscale}
\end{equation}
The scale has units of length squared, so $\Wt$ and $\dE$ have the same units and the field and auxiliary residuals have comparable magnitudes; the term $\ell_2^{-1}$ keeps $s_W$ positive where the magnetic field vanishes. The scaled field equations are
\begin{align}
 \dE+\Amat^{-1}\left[\frac{(\Hop+\Dop+\Smat\Dop)\Wt}{s_W}
 -\vect{w}\times(\nabla_h\times\dE)\right]
 &=\widetilde{\vect{b}}_E,
 \label{eq:fieldrow}\\
 \Wt-s_W\Kop\dE&=\vect{0},
 \label{eq:auxiliaryrow}
\end{align}
with $\widetilde{\vect{b}}_E=\Amat^{-1}(\vect{b}_E-G_P\delta P_e)$.

We write Eqs.~\eqref{eq:fieldrow}--\eqref{eq:auxiliaryrow} as $\mathcal{A}\vect{x}=\vect{b}$, with $\vect{x}=(\dE,\Wt)$, and relax them by damped block-Jacobi sweeps. For smoothing, we follow the defect-correction construction of Ref.~\cite{Chacon2025HMHD} (Sec.~4.1). Each sweep evaluates the residual $\vect{r}=(\vect{r}_E,\vect{r}_W)=\vect{b}-\mathcal{A}\vect{x}$ with the full discrete operators, solves at every node $g$ a local $6\times6$ system $\mathsf{B}_g(\vect{e},\vect{h})_g=\vect{r}_g$, and updates
\begin{equation}
 \vect{x}\leftarrow\vect{x}+\boldsymbol{\Omega}\,\mathsf{B}^{-1}\vect{r},
 \qquad
 \boldsymbol{\Omega}=\operatorname{diag}(\sigma\I,\sigma_W\I),
 \label{eq:jacobi}
\end{equation}
where $\mathsf{B}$ is the block-diagonal matrix of all the $\mathsf{B}_g$ local blocks and $\sigma$ and $\sigma_W$ are damping factors. The iteration matrix is $\I-\boldsymbol{\Omega}\mathsf{B}^{-1}\mathcal{A}$. Because the residual uses the full operators, the fixed point of the sweep solves $\mathcal{A}\vect{x}=\vect{b}$ for any choice of the local block; the block sets only how fast each error component is reduced. The local block, its closed-form inverse, and the weight $\chi_w\in(0,1]$ on its perpendicular curl--curl scale are given in \ref{app:blockjacobi}; the construction follows Ref.~\cite{Chacon2025HMHD} (Appendix~C).

The block-Jacobi smoother is embedded in a geometric multigrid hierarchy with factor-two coarsening and a $V(3,3)$ cycle, with three sweeps before restriction and three after prolongation. Restriction uses tensor-product full weighting and prolongation linear interpolation. The frozen coefficients are injected onto the coarse grids, where Eqs.~\eqref{eq:fieldrow}--\eqref{eq:auxiliaryrow} are rediscretized with the scales of each level, and identity rows remain identity rows on every level. The coarsest level is treated by eight smoothing sweeps rather than by a direct solve. The sweeps remove the grid-scale error, where the Hall stiffness is largest, and the coarse levels remove the smooth error.

Each preconditioner application first solves the pressure block and then solves Eqs.~\eqref{eq:fieldrow}--\eqref{eq:auxiliaryrow} with GMRES, right-preconditioned by one $V$-cycle per iteration, until the residual falls by a prescribed relative tolerance or an iteration cap is reached; Section~\ref{sec:verification} gives the values. The number of inner iterations depends on the right-hand side, so the preconditioner changes across outer iterations, necessitating flexible GMRES (FGMRES)~\cite{Saad1993FGMRES} as the outer Krylov solver, for which we use PETSc~\cite{Balay1997PETSc}.

\subsection{Analytical mass matrix and accelerated quasi-Newton solve}
\label{subsec:fluid}

Each Jacobian action needs the ion current response to a push-field update. With the midpoint positions $\bar{\vect{x}}_p$ and the gathered magnetic field $\B_p$ frozen at the Newton iterate, Eq.~\eqref{eq:push} gives a midpoint velocity that is affine in the gathered push field $\E^*_p$,
\begin{equation}
 \begin{gathered}
 \bar{\vect{v}}_p=\mathsf{R}_p\left(\vect{v}_p^n+\alpha_p\E^*_p\right),
 \qquad
 \alpha_p=\frac{q_p}{m_p}\frac{\Delta t}{2},\\
 \mathsf{R}_p=\frac{\I+\vect{b}_p\vect{b}_p^T-[\vect{b}_p]_\times}{1+b_p^2},
 \qquad \vect{b}_p=\alpha_p\B_p,
 \end{gathered}
 \label{eq:vbar}
\end{equation}
with $[\vect{b}]_\times\vect{u}=\vect{b}\times\vect{u}$. A push-field update $\delta\E^*$ changes the gathered field, Eq.~\eqref{eq:gather}, by $\delta\E^*_p=\sum_{g'}S(\bar{\vect{x}}_p-\vect{x}_{g'})(\mathsf{F}\delta\E^*)_{g'}$ and the midpoint velocity by $\delta\bar{\vect{v}}_p=\alpha_p\mathsf{R}_p\,\delta\E^*_p$. Depositing this change with Eq.~\eqref{eq:deposit} gives $\delta\Ji=\mathsf{F}\Mmat\mathsf{F}\,\delta\E^*$, where the mass matrix $\Mmat$ has the $3\times3$ blocks
\begin{equation}
 \Mmat_{gg'}=\frac{1}{\Delta V_g}\sum_p q_p\alpha_p\mathsf{R}_p\,
 S(\bar{\vect{x}}_p-\vect{x}_g)\,S(\bar{\vect{x}}_p-\vect{x}_{g'}).
 \label{eq:massmatrix}
\end{equation}
The density responds through the continuity equation,
\begin{equation}
 \delta\rhoc\simeq-\frac{\Delta t}{2}\nabla_h\cdot\delta\Ji,
 \label{eq:densityresponse}
\end{equation}
evaluated from the ion-current update at nodes above a density threshold. Equation~\eqref{eq:densityresponse} is not the exact derivative of the particle deposition, but it is conservative. The terms omitted by freezing the orbits, the position response and $\bar{\vect{v}}_p\times\delta\B$, remain in the nonlinear residual, which advances the particles at every Newton iterate~\cite{Angus2023,Kumar2026ThetaImplicitHybrid}.

A direct computation of the mass matrices~\cite{Lapenta2017,Angus2023} requires evaluating Eq.~\eqref{eq:massmatrix} by a particle loop at each Newton iterate and storing one $3\times3$ block for each pair of nodes within the shape-function support, $9(2s+1)^d$ values per node, or 243 in three dimensions with linear shapes. 
To avoid this complexity, the mass matrices can instead be approximated analytically, extending the local-node approximation of the preconditioner to the full stencil \cite{Chen2015Darwin}. Replacing the particle sum of each ion species in Eq.~\eqref{eq:massmatrix} by its average over a smooth density, $\sum_pw_pf(\bar{\vect{x}}_p)\to\int n(\bar{\vect{x}})f(\bar{\vect{x}})\,d\bar{\vect{x}}$, makes every block of $\Mmat$ a local function of the density and magnetic field, as $\boldsymbol{\Sigma}_g$ is in Eq.~\eqref{eq:Sdef}. If the density and magnetic field vary little over the shape-function support, this gives the analytical mass matrix (AMM) approximation:
\begin{equation}
 \begin{aligned}
 \Mmat_{gg'}&\simeq\sum_{\rm species}\alpha\,\rho_g\,\mathsf{R}(\vect{b}_g)\,K_{g-g'},\\
 K_{g-g'}&=\frac{1}{\Delta V}\int S(\bar{\vect{x}}-\vect{x}_g)\,S(\bar{\vect{x}}-\vect{x}_{g'})\,d\bar{\vect{x}}
 =\prod_dS_{2s+1}(g_d-g'_d),
 \end{aligned}
 \label{eq:fluidM}
\end{equation}
where the sum runs over the ion species, $\rho_g$ is the deposited charge density of each species at the half step, $\vect{b}_g=\alpha\B_g$, and $S_{2s+1}$ is the centered B-spline of degree $2s+1$, the autocorrelation of the shape function; for linear shapes, $K$ has the weights $(1/6,2/3,1/6)$ in each direction. The kernel is computed once, and the nodal coefficient, four values per node and species, is computed from the grid charge density and the total magnetic field once per Newton iteration. Each Jacobian action then applies a $(2s+1)^d$ stencil and a $3\times3$ product per node, with no particle loop, and the resulting current is filtered, summed across box boundaries, and reflected at walls exactly as the particle deposit. Because the rows of $K$ sum to one, the response reduces to the cold-fluid response $\sum_{\rm species}\alpha\rho\mathsf{R}\,\delta\E^*$ for a smooth push field. In cylindrical geometry, restricted to the azimuthally symmetric mode, the radial weights become $\frac12(i+i')S_{2s+1}(i-i')$ in units of $\Delta r$, with tabulated one-sided weights next to the axis and the outer wall.

The analytical response differs from Eq.~\eqref{eq:massmatrix} by the particle noise, the density variation within the stencil, and the spread of the magnetic field among the particles. This difference is confined to the Jacobian action, where it enters the field equations through the ion part of the Hall term, whose size relative to the identity is set by $b=\Omega_{ci}\Delta t/2$; it therefore changes the rate of Newton convergence but not the converged solution.

With the AMM in $\widetilde{\mathsf{J}}_k$ and the preconditioner of Sections~\ref{subsec:pcapprox} and~\ref{subsec:fieldblock}, the particles are advanced only to evaluate the residual, once per Newton iteration and line-search trial, while every Krylov iteration applies $\widetilde{\mathsf{J}}_k$ and $\mathsf{P}^{-1}$, both mesh operations. Algorithm~\ref{alg:quasinewton} summarizes one time step.


\begin{algorithm}[!tb]
\small\linespread{1}\selectfont
\caption{Accelerated quasi-Newton advance of one $\theta$-implicit hybrid-PIC time step.}
\label{alg:quasinewton}
\begin{enumerate}\setlength{\itemsep}{2pt}\setlength{\parskip}{0pt}
 \item Begin with the initial guess $\vect{U}_0=(\E^{n},P_e^{n})$ from time $t_n$, and set $k=0$.
 \item Evaluate $\vect{F}_k$: advance every particle to $\bar{\vect{x}}_p$ and $\bar{\vect{v}}_p$ by a per-particle Picard iteration of Eq.~\eqref{eq:push} with the push field and magnetic field of $\vect{U}_k$, deposit $\rhoc$ and $\Ji$, and evaluate Eqs.~\eqref{eq:ohm} and~\eqref{eq:pe}.
 \item If $\|\vect{F}_k\|\leq\epsilon_t$, go to step 7.
 \item Build the mesh operators at $\vect{U}_k$: the nodal coefficients of the AMM response, Eq.~\eqref{eq:fluidM}, and the preconditioner coefficients of Eqs.~\eqref{eq:Kandw}--\eqref{eq:Sdef} and~\eqref{eq:pressureop}.
 \item Solve the approximate Jacobian system in Eq.~\eqref{eq:newton} with FGMRES to the relative tolerance $\zeta_k$ of Eq.~\eqref{eq:forcing}. Each iteration applies $\widetilde{\mathsf{J}}_k$, evaluating the field equations with the ion current $\Ji+\delta\Ji$ and density $\rhoc+\delta\rhoc$ of Eqs.~\eqref{eq:fluidM} and~\eqref{eq:densityresponse} and no particle advance, and $\mathsf{P}^{-1}$: the pressure block, then inner GMRES on the field block preconditioned by a multigrid $V$-cycle.
 \item Update $\vect{U}_{k+1}=\vect{U}_k+\lambda_k\delta\vect{U}_k$ with the line search of Section~\ref{subsec:newton}. Each trial step evaluates the residual as in step 2, and the residual of the accepted step is $\vect{F}_{k+1}$. Set $k\leftarrow k+1$ and go to step 3.
 \item Complete the time step: advance $\B$ to $n+1$ with Eq.~\eqref{eq:faraday}, extrapolate $P_e$ to $n+1$, and advance the particles with $\vect{x}_p^{n+1}=2\bar{\vect{x}}_p-\vect{x}_p^n$ and $\vect{v}_p^{n+1}=2\bar{\vect{v}}_p-\vect{v}_p^n$.
\end{enumerate}
\end{algorithm}

\section{Verification and solver characterization}
\label{sec:verification}

\subsection{Linear wave propagation}
\label{subsec:waves}

The first two linear tests adapt the standing-wave problems of Ref.~\cite{Chacon2025HMHD} (Secs.~5.1--5.2) from Hall MHD to kinetic ions in the cold-ion limit: a parallel whistler wave, which exercises the stiff Hall response that the preconditioner targets, and an oblique kinetic Alfvén wave, which adds the electron-pressure coupling. Both use periodic collocated meshes, linear particle shapes, $\theta=1/2$, $\eta=\eta_h=0$, and the isothermal closure $P_e=n_eT_e$ ($\gamma=1$), which removes the pressure unknown and, with it, the pressure block of the preconditioner. The physics-based preconditioner is used throughout, with relative Newton and Krylov tolerances of $10^{-6}$. Because we use these tests to measure discretization errors, we fix the Krylov tolerance at this value rather than setting it by the adaptive forcing of Eq.~\eqref{eq:forcing}, which we use for the GEM problem (Section~\ref{subsec:gem}).  Unless stated otherwise, the preconditioner uses the damping factors $\sigma=1.3$ and $\sigma_W=1.2$, $\chi_w=0.4$, and inner GMRES with relative tolerance $0.3$ and at most eight iterations; the coarsest multigrid level is given with each test. The whistler test uses the quasi-Newton Jacobian with deposited mass matrices and the oblique test the JFNK Jacobian (Section~\ref{subsec:newton}), so that both Jacobian actions are exercised; every step converged in one or two Newton iterations. We define $v_A=B_0/\sqrt{\mu_0n_0m_i}$, $\tau_A=L/v_A$, $d_i=v_A/\Omega_{ci}$, and $\beta_e=2\mu_0n_0T_e/B_0^2$, with $\B_0=B_0\hat{\vect{z}}$.

The reference for each test is the exact solution of the linearized hybrid equations, with cold ions and isothermal electrons, for the same initial perturbation. We obtain the simulated frequency $\omega_{\rm sim}$ by fitting $A|\sin\omega t|$ to the amplitude of the initialized Fourier mode of $B_y$ (whistler), or $A|\cos\omega t|$ to that of $u_{ix}$ (oblique wave), and compare it with the frequency $\omega_{\rm ref}$ of the same wave in the reference:
\begin{equation}
 \varepsilon_\omega=\frac{\omega_{\rm ref}-\omega_{\rm sim}}{\omega_{\rm ref}}\simeq a\,\Delta x^2+b\,\Delta t^2.
 \label{eq:freqerror}
\end{equation}
For each test, the constants $a$ and $b$ come from a least-squares fit to all of its runs. Table~\ref{tab:waves} lists $\varepsilon_\omega$ and its spatial and temporal parts, $a\Delta x^2$ and $b\Delta t^2$; their sum reproduces every measured $\varepsilon_\omega$ to within $10^{-5}$. A third test, the normal modes of a plasma-filled conducting cylinder, exercises the cylindrical mesh and its boundary conditions (Section~\ref{subsec:rzmodes}). 

\begin{table}[!tb]
 \centering
 \small
 \caption{Frequency error $\varepsilon_\omega$, Eq.~\eqref{eq:freqerror}, of the linear wave tests and its spatial and temporal parts. The last row of each test is the run shown in Fig.~\ref{fig:whistler} or~\ref{fig:kaw}. Meshes are $N_z$ for the whistler and $N_x\times N_z$ for the oblique wave.}
 \label{tab:waves}
 \setlength{\tabcolsep}{4pt}
 \begin{tabular}{llrrrrr}
  \toprule
  & & & & \multicolumn{3}{c}{Frequency error (\%)} \\
  \cmidrule(lr){5-7}
    Test & Mesh & $\Delta t/\tau_A$ & Particles/cell & $\varepsilon_\omega$ & $a\Delta x^2$ & $b\Delta t^2$ \\
  \midrule
  Whistler & 64 & $10^{-4}$ & 16\,384 & 0.337 & 0.325 & 0.0125 \\
           & 64 & $10^{-4}$ & 65\,536 & 0.339 & 0.325 & 0.0125 \\
           & 128 & $10^{-4}$ & 16\,384 & 0.0938 & 0.0813 & 0.0125 \\
           & 128 & $5\times10^{-5}$ & 32\,768 & 0.0846 & 0.0813 & 0.00313 \\
           & 128 & $5\times10^{-5}$ & 100\,000 & 0.0848 & 0.0813 & 0.00313 \\
  \midrule
  Oblique  & $256\times64$ & $4\times10^{-5}$ & 256 & 1.01 & 0.944 & 0.0656 \\
           & $256\times64$ & $4\times10^{-5}$ & 1\,024 & 1.01 & 0.944 & 0.0656 \\
           & $256\times64$ & $2\times10^{-5}$ & 256 & 0.961 & 0.944 & 0.0164 \\
           & $512\times128$ & $4\times10^{-5}$ & 256 & 0.302 & 0.236 & 0.0656 \\
           & $512\times128$ & $2\times10^{-5}$ & 256 & 0.253 & 0.236 & 0.0164 \\
           & $1024\times256$ & $10^{-5}$ & 64 & 0.0634 & 0.0590 & 0.00410 \\
  \bottomrule
 \end{tabular}
\end{table}

\subsubsection{Parallel whistler wave}
\label{subsec:whistler}

\begin{figure*}[!t]
 \centering
 \includegraphics[width=\textwidth]{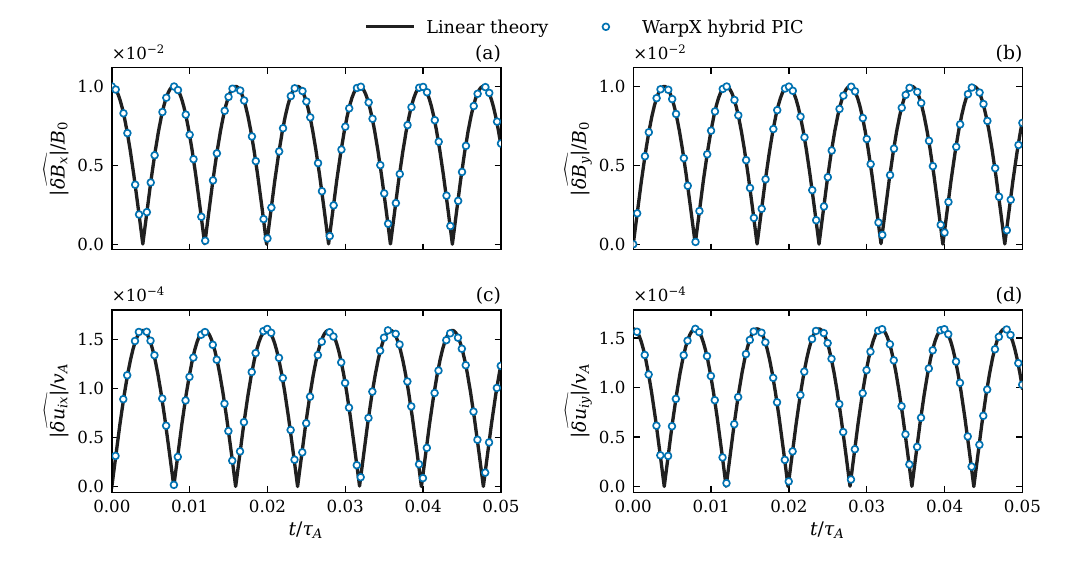}
 \caption{Parallel whistler verification: Fourier amplitudes of the transverse magnetic field and ion velocity. Open circles show the hybrid-PIC solution; solid lines show the unfitted cold-ion linear reference.}
 \label{fig:whistler}
\end{figure*}

In a one-dimensional domain of length $L$, $d_i/L=10$ and $k=2\pi/L$. The initial perturbation is
\begin{equation}
 \frac{\delta B_x}{B_0}=\epsilon\cos(kz),\qquad
 \frac{\delta u_{iy}}{v_A}=\frac{\epsilon}{kd_i}\sin(kz),
\end{equation}
with $\epsilon=10^{-2}$, and the ions are cold Maxwellian with $v_{thi}=0.005v_A$ and $\beta_e=10^{-4}$. The whistler frequency is
\begin{equation}
 \omega_w=\frac{kv_A}{2}\left[kd_i+\sqrt{(kd_i)^2+4}\right],
 \label{eq:whistler-frequency}
\end{equation}
which approaches $d_ik^2v_A$ at large $kd_i$. Figure~\ref{fig:whistler} compares the run with 128 cells, $\Delta t=5\times10^{-5}\tau_A$, and $10^5$ particles per cell over $10^3$ steps, with the reference. The large particle count is needed because the ion-velocity response is a factor $kd_i\approx63$ weaker than the magnetic one. The frequency error is $0.085\%$, and the peak-normalized root-mean-square waveform errors are below $1.2\%$ for all four components. A fourfold increase in particles per cell changes the error by less than $2\times10^{-5}$, so the error is set by the mesh and the time step, and the spatial part dominates in every run (Table~\ref{tab:waves}). Both parts agree with the leading-order dispersion errors of the scheme: at 64 cells, the fitted spatial part, $0.325\%$, compares with $(k\Delta z)^2/3=0.321\%$ for the centered curl composition, and at $\Delta t=10^{-4}\tau_A$, the fitted temporal part, $0.0125\%$, compares with $(\omega_w\Delta t)^2/12=0.013\%$ for the implicit midpoint rule.

\subsubsection{Oblique kinetic Alfvén wave}
\label{subsec:kaw}

\begin{figure*}[!t]
 \centering
 \includegraphics[width=\textwidth]{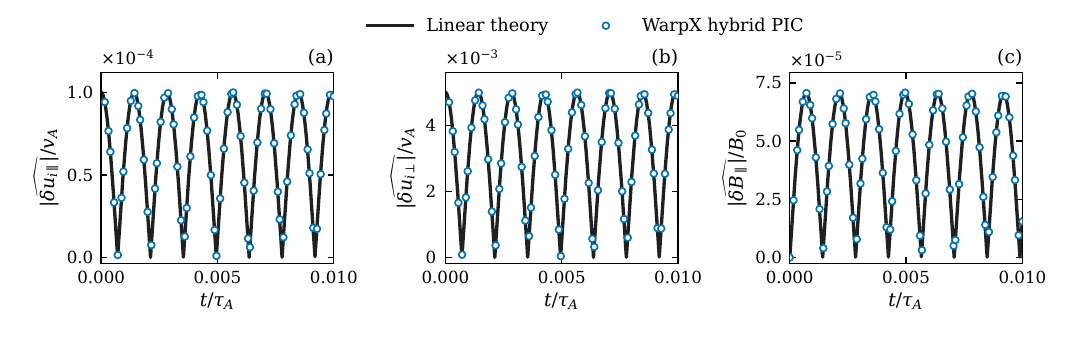}
 \caption{Oblique kinetic Alfvén wave verification following the benchmark of Ref.~\cite{Chacon2025HMHD}, adapted to cold kinetic ions. Panels show parallel and perpendicular ion-velocity amplitudes and the parallel magnetic-field amplitude. Symbols and lines have the same meaning as in Fig.~\ref{fig:whistler}.}
 \label{fig:kaw}
\end{figure*}

The second test follows the kinetic Alfvén-wave benchmark of Ref.~\cite{Chacon2025HMHD}. The domain has length $L_z=10L$ along $\B_0$ and $L_x=L$ across it, with $L=\pi d_i/5$. The perturbation has one wavelength along $\B_0$ and five across it: $k_\parallel=2\pi/L_z$ and $k_\perp=10\pi/L$, so that $k_\parallel d_i=1$ and $k_\perp d_i=50$. To leading order, $\omega\simeq\rho_sk_\parallel k_\perp v_A$ with $\rho_s=\sqrt{T_e/m_i}/\Omega_{ci}$, well above $\Omega_{ci}$ at these parameters. The equilibrium pressure is placed in the electrons, $\beta_e=10^4$ with $v_{thi}=10^{-3}v_A$, which keeps the ions cold while preserving $\rho_s/d_i=\sqrt{\beta/2}$ as in the fluid benchmark. The initial perturbations are $\delta u_{i\parallel}/v_A=\epsilon\cos\phi$ and $\delta u_{i\perp}/v_A=50\epsilon\cos\phi$, with $\phi=k_\perp x+k_\parallel z$, $u_{i\parallel}=u_{iz}$, $u_{i\perp}=u_{ix}$, and $\epsilon=10^{-4}$; density and magnetic field are unperturbed. Particles start on a regular lattice, and one binomial-filter pass per direction is applied to the deposited moments and the gathered field. Figure~\ref{fig:kaw} compares the run with $1024\times256$ cells, 64 particles per cell, and $\Delta t=10^{-5}\tau_A$ through $t=0.01\tau_A$ with the reference; the frequency error is $0.063\%$, and the peak-normalized root-mean-square errors are below $0.7\%$ for all three components. As in the whistler test, the two-term fit reproduces every run, the spatial part dominates on every mesh, and a fourfold increase in particles per cell leaves the error unchanged (Table~\ref{tab:waves}).

\subsubsection{Cylindrical normal modes}
\label{subsec:rzmodes}

The third linear test considers cylindrical $(r,z)$ geometry, restricted to the azimuthally symmetric mode, in a bounded domain with the boundary conditions of Section~\ref{subsec:pcapprox}. It follows the hybrid-PIC normal-mode test of Ref.~\cite{Groenewald2023}: a uniform plasma fills a conducting cylinder of radius $R=51.2d_i$ and length $L_z=204.8d_i$, periodic in $z$, with $B_0=0.5$~T along the axis, $n_0=2.4\times10^{20}\,$m$^{-3}$, ions at $T_i=25.5$~eV ($\beta_i=0.01$), cold electrons, and $\eta=5\times10^{-4}\,\Omega\,$m. Thermal noise excites the normal modes of the cylinder. The azimuthal electric field is decomposed into radial Bessel modes, through a Hankel transform on the zeros of $J_1$, and into Fourier modes in $z$ and $t$, and the spectra are compared with the cold-plasma dispersion of the fast (compressional) and slow (shear) branches of a conducting cylinder~\cite{Stix1992}, which follows from $B_0$, $n_0$, and $R$ alone. The mesh is $128\times512$ cells ($\Delta r=\Delta z=0.4d_i$) with 2\,000 particles per cell and linear particle shapes. The outer wall is a perfect conductor with specularly reflecting ions, and the axis is a symmetry boundary. The time step is $\Delta t=0.08\,\tau_{ci}$, with $\tau_{ci}=2\pi/\Omega_{ci}$, and the run covers $200\tau_{ci}$ with $\theta=1/2$, the AMM, the preconditioner with a coarsest level of 24 cells per direction, a Newton relative tolerance of $10^{-6}$, and the adaptive forcing of Eq.~\eqref{eq:forcing}.

\begin{figure*}[!t]
 \centering
 \includegraphics[width=\textwidth]{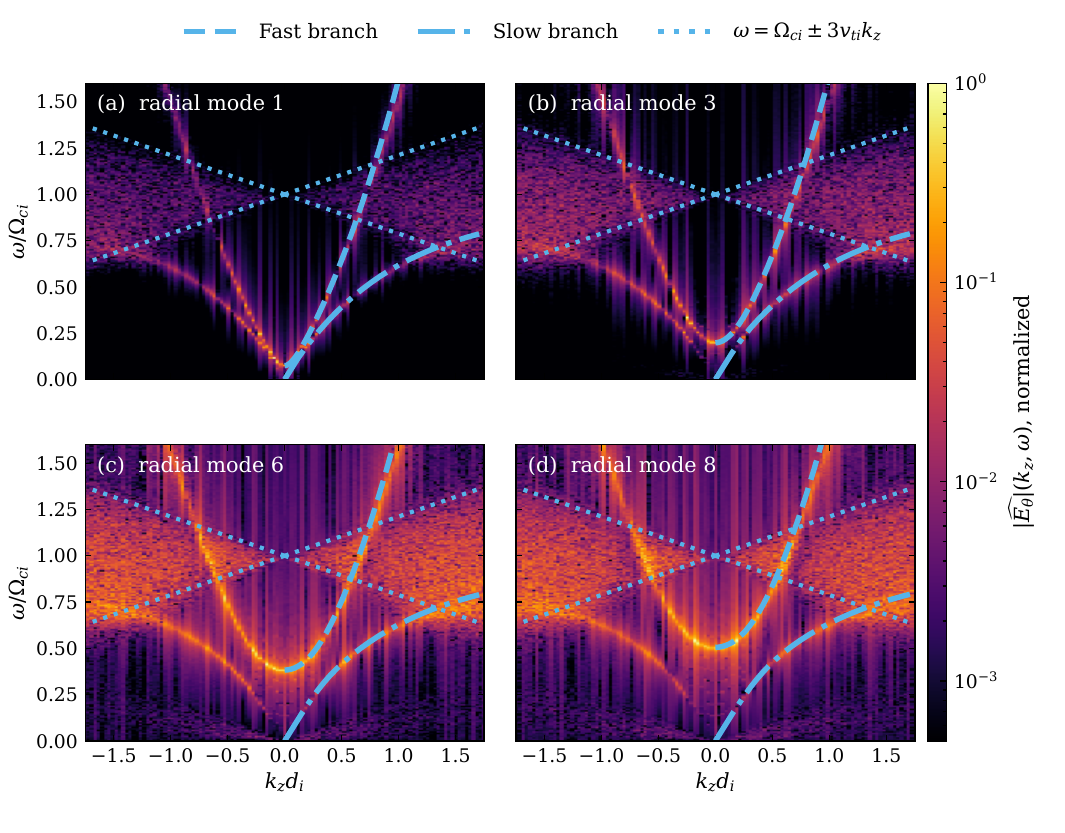}
 \caption{Cylindrical normal modes ($m=0$): power spectrum of the azimuthal electric field for radial modes 1, 3, 6, and 8 versus axial wavenumber and frequency. Dashed and dash-dotted lines show the cold-plasma fast and slow branches of a conducting cylinder; dotted lines show the ion-cyclotron thermal band $\omega=\Omega_{ci}\pm3v_{ti}k_z$.}
 \label{fig:rzmodes}
\end{figure*}

Figure~\ref{fig:rzmodes} shows the spectra of the radial modes 1, 3, 6, and 8. The spectra follow the fast branch across the resolved wavenumber range and the slow branch below the ion-cyclotron frequency, with the thermal broadening $\omega=\Omega_{ci}\pm3v_{ti}k_z$ visible around $\Omega_{ci}$, as in Ref.~\cite{Groenewald2023}. At this time step, the grid-scale whistler stiffness is $c_w\ell_2=12.6$, with $c_w$ from Eq.~\eqref{eq:Kandw} and the cylindrical $\ell_2$. Every step converged in two or three Newton iterations, with an average of 7.0 FGMRES iterations per Newton iteration and 1.5~s per step on one NVIDIA RTX 6000 Ada GPU.

\subsection{GEM magnetic reconnection}
\label{subsec:gem}

\begin{figure*}[!t]
 \centering
 \includegraphics[width=\textwidth]{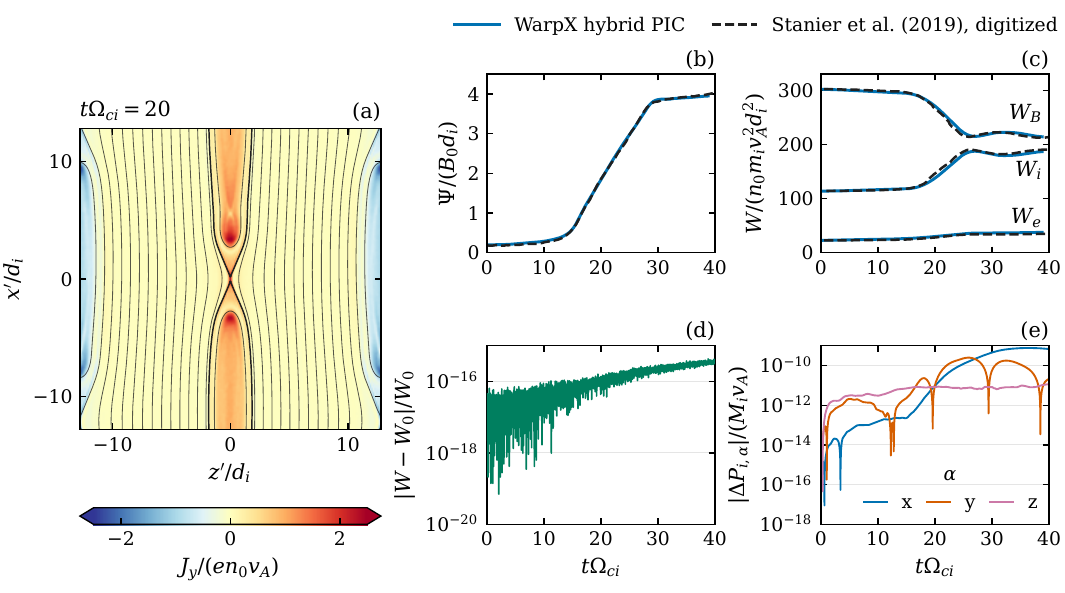}
\caption{GEM reconnection in the setup of Ref.~\cite{Stanier2019}, with 320 particles per cell, $\Delta t\Omega_{ci}=0.01$, and Newton relative tolerance $10^{-13}$. (a) Out-of-plane current at $t\Omega_{ci}=20$ with flux contours (b) reconnected flux (c) magnetic, ion kinetic, and electron internal energies in units of $n_0m_iv_A^2d_i^2$; (d) relative total-energy error; (e) change of the total ion momentum components in units of $M_iv_A$. In (b,c), our curves are shifted earlier by $1.2\,\Omega_{ci}^{-1}$ to align the reconnection onset with the dashed curves, digitized from Ref.~\cite{Stanier2019}.}

 \label{fig:gem}
\end{figure*}

We consider the doubly periodic GEM reconnection problem of Stanier et al.~\cite{Stanier2019}, whose closed domain allows the conservation of total energy and momentum to be measured directly. Two oppositely directed Harris sheets of half-width $\lambda=0.5d_i$ lie at $z=\pm L_z/4$ in a square domain with $L_x=L_z=25.6d_i$. The background density is $0.2n_0$, the guide field and initial ion drift are zero, and $T_i/T_e=5$, with reference beta $2\mu_0n_0(T_i+T_e)/B_0^2=0.998$. Reconnection is seeded by
\begin{equation}
 \delta A_y=0.1B_0d_i\cos(2\pi x/L_x)\cos(4\pi z/L_z).
\end{equation}
The mesh has $384\times384$ collocated cells. The electron pressure starts at uniform temperature and evolves with $\gamma=5/3$ and dissipative heating, without conduction ($\kappa_e=0$), as in Ref.~\cite{Stanier2019}; in the ion units defined above,
\begin{equation}
 \frac{\eta}{\mu_0v_Ad_i}\simeq4.99\times10^{-4},\qquad
 \frac{\eta_h}{\mu_0v_Ad_i^3}\simeq2.00\times10^{-4}.
 \label{eq:gem-dissipation}
\end{equation}
All GEM runs use linear particle shapes, $\theta=1/2$, two binomial-filter passes on the deposited moments and the gathered fields, van Albada pressure advection with the velocity bound of Section~\ref{subsec:model}, and the line search of Section~\ref{subsec:newton}. The run of Fig.~\ref{fig:gem} follows the reference setup, with 213 sheet and 107 background particles per cell (320 on average, as in the reference) and $\Delta t\Omega_{ci}=0.01$. It uses the AMM and a Newton relative tolerance of $10^{-13}$.

Figure~\ref{fig:gem} shows the out-of-plane current at $t\Omega_{ci}=20$, the reconnected flux, the energies, and the conservation errors through $t\Omega_{ci}=40$. Reconnection sets in about $1.2\,\Omega_{ci}^{-1}$ later than in the reference. With our curves shifted earlier by this time, as in Fig.~\ref{fig:gem}(b,c), the reconnected flux follows the reference from onset to saturation, and the magnetic and ion energies follow it, with the energy transfer lagging slightly during the fast phase. The magnetic energy released by $t\Omega_{ci}=40$ agrees with the reference to within $0.6\%$. The onset time depends on the particle noise that seeds the tearing mode: halving the number of particles per cell advances the onset by $0.7\,\Omega_{ci}^{-1}$, and quadrupling it delays the onset by $0.25\,\Omega_{ci}^{-1}$, while the reconnection rate, the saturated flux, and the final magnetic energy change by less than $1\%$ (not shown). The relative total-energy error remains below $5\times10^{-16}$, at the level of floating-point roundoff, and every component of the total ion momentum changes by less than $10^{-9}M_iv_A$, where $M_i$ is the total ion mass.

\subsection{Solver performance}
\label{sec:performance}

\begin{table}[!tb]
 \centering
 \small
 \caption{GEM solver cost over the first 25 steps at $\Delta t\Omega_{ci}=0.05$ with $2.3\times10^8$ particles, Newton relative tolerance $10^{-6}$, and particle tolerance $10^{-13}$, on one NVIDIA RTX 6000 Ada GPU. ``Ion response'' is the time spent on the ion current response in the Jacobian-vector products, and includes particle re-advance (in JFNK, estimated) and deposition and application of the mass matrices (either particle-based, MM, or analytical, AMM). ``Preconditioner'' accounts for the time spent applying $\mathsf{P}^{-1}$.}
 \label{tab:gem-cost}
 \setlength{\tabcolsep}{4pt}
 \begin{tabular}{lrrrr}
  \toprule
  & JFNK & JFNK + PC & MM + PC & AMM + PC \\
  \midrule
  \# FGMRES / Newton iteration & 357 & 6.46 & 6.35 & 6.35 \\
  \# Newton iterations / step & 2.28 & 2.00 & 2.04 & 2.04 \\
  Ion response (s) & $\approx$6\,983 & $\approx$111 & 35.3 & 0.03 \\
  Preconditioner (s) & -- & 6.8 & 6.9 & 7.0 \\
  Wall-clock time (s) & 7\,094 & 149 & 73.3 & 38.1 \\
  Speed-up & 1 & 48 & 97 & 186 \\
  \bottomrule
 \end{tabular}
\end{table}

Table~\ref{tab:gem-cost} compares four solver configurations over the first 25 steps of the GEM problem. The runs share the executable, the initial particles, and all other settings. In standard JFNK, the Krylov solve takes 357 FGMRES iterations per Newton iteration, with each iteration re-advancing the particles. The preconditioner reduces this count to 6.5 and the time per step 48-fold. Particle-based mass matrices replace the particle re-advance in the Jacobian actions by a mesh operation, but depositing and applying them still takes a significant amount of time (35~s of the 73~s of wall-clock time). Instead, the AMM computes the ion response in 0.03~s without increasing iteration counts. Combining AMM with the preconditioner decreases the wall-clock time by 186-fold relative to standard JFNK. The remaining time is dominated by the nonlinear particle advance and the preconditioner application.

\begin{figure*}[!t]
 \centering
 \includegraphics[width=\textwidth]{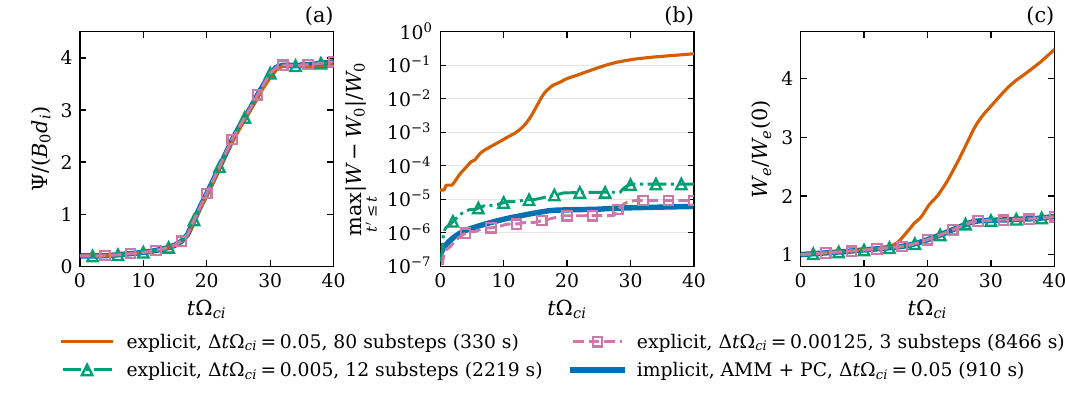}
 \caption{GEM reconnection through $t\Omega_{ci}=40$ with $2.3\times10^8$ particles, computed with the implicit solver (AMM and preconditioner) at $\Delta t\Omega_{ci}=0.05$, with Newton relative tolerance $10^{-3}$, particle tolerance $10^{-4}$, and fixed Krylov tolerance $10^{-2}$, and with the explicit hybrid-PIC solver at the same time step and at ten- and forty-times-smaller steps. The number of explicit field substeps is the whistler stability limit. (a) Reconnected flux; (b) largest relative total-energy error up to time $t$; (c) electron internal energy. The legend gives the wall-clock times on one NVIDIA RTX 6000 Ada GPU.}
 \label{fig:gem-explicit}
\end{figure*}

Figure~\ref{fig:gem-explicit} compares GEM runs through $t\Omega_{ci}=40$ with the explicit hybrid-PIC solver of WarpX~\cite{Groenewald2023}, which uses the same physical model and subcycles the field advance with fourth-order Runge--Kutta steps to satisfy the whistler stability limit. The implicit run, with the looser tolerances given in the caption, completes in 910~s and keeps the total-energy error below $6.2\times10^{-6}$. The explicit solver with the same time step (which results in 80 field substeps due to the whistler stability constraint) completes in 330~s, 2.8 times faster than implicit, but results in $\mathcal{O}(1)$ errors in total energy conservation, more than four orders of magnitude larger than the implicit energy conservation error. With 120 field substeps (not shown) the result is unchanged, so the energy conservation error originates in the explicit particle step rather than the field advance. Reducing the explicit particle timestep closes the conservation-error gap. At a ten-times-smaller particle timestep, $\Delta t\Omega_{ci}=0.005$, the explicit run reproduces the reconnected flux and electron heating of the implicit run, but takes 2219~s, 2.4 times longer than implicit, and its energy conservation error remains $\sim4.6$ times larger. With a forty-times smaller particle timestep, $\Delta t\Omega_{ci}=0.00125$, the explicit energy conservation error becomes comparable to the implicit one but the explicit run takes 8466~s, 9.3 times longer than implicit.

\section{Three-dimensional FRC formation}
\label{sec:frc}

\begin{figure*}[!t]
 \centering
 \includegraphics[width=\textwidth]{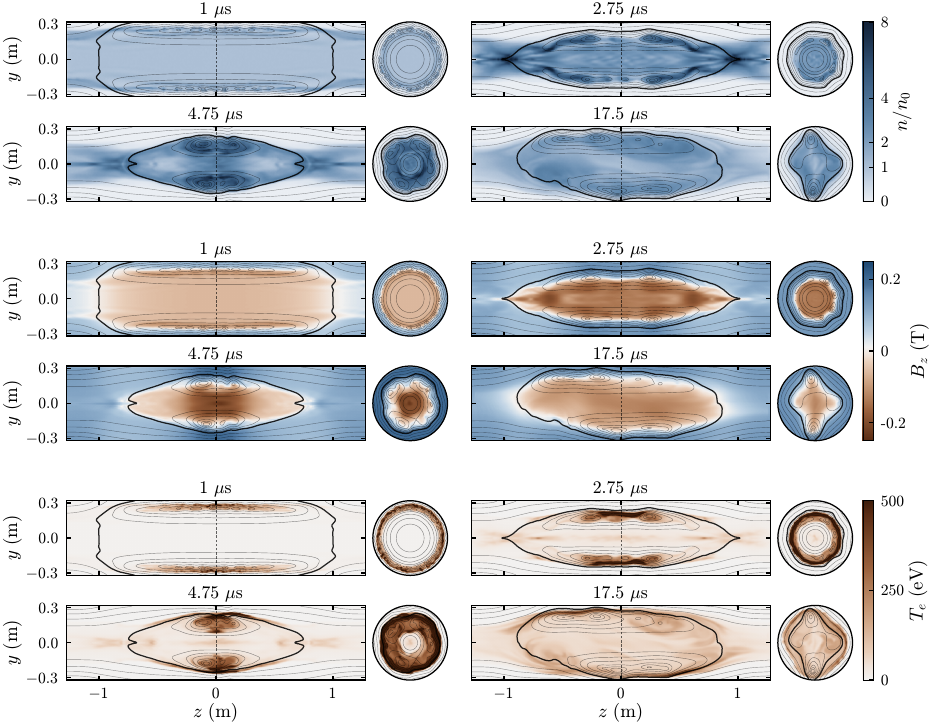}
 \caption{Cross sections of the density $n$, the axial magnetic field $B_z$, and the electron temperature $T_e$ at four times for the 3D FRC formation problem. The vertical dashed line on the $z$-$y$ contours is the plane of the polar slice. The white and black solid lines overlaid on the contours show the poloidal flux surfaces, with the thick line the separatrix.}
 \label{fig:frc}
\end{figure*}

As a final demonstration, we simulate the formation of a field-reversed configuration (FRC)
in a three-dimensional field-reversed theta-pinch discharge. The problem combines rapidly
reversing applied fields, a vacuum region that opens between the plasma and the wall, and
magnetic reconnection on the reversing surface and at the ends of the plasma column. The setup
follows the higher-bias discharges of Ref.~\cite{Omelchenko2015}, but instead of assuming an
adiabatic electron temperature, we evolve the electron pressure with Eq.~\eqref{eq:pe},
including Ohmic heating and electron--ion energy exchange, and include ion--ion collisions;
electron thermal conduction is not included. Converging flows on a Cartesian grid acquire
spurious perturbations where curved fronts cross cell boundaries~\cite{Joggerst2014}, and
because the grid is unchanged by a $90^\circ$ rotation, this error appears as an $m=4$
azimuthal imprint~\cite{Beresnyak2022}. To suppress it, the whole solution is rotated rigidly
about the axis by $\sim 34.38^\circ$, an irrational number of degrees, every ten time
steps. The solution therefore never returns to the same orientation relative to the grid and
samples all orientations nearly uniformly, so grid-locked error stays bounded instead of
accumulating. The ions are rotated exactly and the fields are remapped with quintic splines; each remap leaves the magnetic flux through every axial plane unchanged.

A uniform deuterium plasma with $n_0 = 5 \times 10^{19}$~m$^{-3}$ and $T_i = T_e = 5$~eV
fills a conducting cylinder of radius $0.32$~m and length $2.56$~m, threaded by a bias field
of $-0.06$~T. A main coil and two end coils, prescribed through the vector-potential
splitting of Section~\ref{subsec:model}, reverse the field at the wall over a quarter
period of $4~\mu$s and form magnetic mirrors at the ends. As in Ref.~\cite{Omelchenko2015}, 
ion–ion Coulomb collisions are included, here with a pairwise binary-collision model. 
The resistivity is the Spitzer
value without the anomalous contribution of Ref.~\cite{Omelchenko2015}. 
The mesh has $128 \times 128 \times 480$ cells with $8.4 \times 10^8$ ions, and the
discharge is followed to $20~\mu$s with a time step of $\Delta t\,\Omega_{ci} = 0.016$, about
$440$ times the explicit whistler limit. The simulation takes $6098$ time steps and $12.7$~h of wall-clock time on $20$ NVIDIA A100 GPUs
(five Perlmutter nodes). With a relative Newton tolerance of
$10^{-4}$, a step needs on average $2.4$ Newton iterations and $7.5$ preconditioned FGMRES iterations per Newton iteration.

Figure~\ref{fig:frc} shows the formation. The imploding current sheath first heats a thin
layer at the wall while the bias field remains trapped in the column. By $2.5~\mu$s the
field lines have reconnected near the end coils, closing the reversed flux into an elongated
configuration bounded by X-points, and about half of the chamber has emptied into a
near-vacuum region. The tension of the reconnected field lines then drives an axial
contraction, after which the configuration rebounds and relaxes into an elongated FRC. 
This sequence is consistent with Ref.~\cite{Omelchenko2015}, including disruption due to the tilt instability, which is expected as the formed FRC exceeds the typical $S^*/E\lesssim 3.7$ MHD stability bound~\cite{Steinhauer2011}.

The main difference from the present results compared to~\cite{Omelchenko2015} is magnitude of the electron temperature. Within the separatrix, the
density-weighted $T_e$ settles near $140$~eV, about half of the ion temperature, whereas the
adiabatic law applied to the same density gives about 22~eV, consistent with the peak of
$25$~eV in Ref.~\cite{Omelchenko2015}. The additional heating comes from Ohmic dissipation in
the imploding sheath and at the reconnection sites, and subsequent flows within the separatrix then
redistribute this energy into a broad $T_e$ profile.

These results show that the implicit solver captures the three-dimensional formation dynamics,
including vacuum regions, time-varying applied fields, and reconnection, at a single time step
far above the explicit whistler limit. The formation process agrees qualitatively with
Ref.~\cite{Omelchenko2015}, while the evolved electron energy yields a substantially hotter
electron population than the adiabatic model. A quantitative study of the formed
configuration and its sensitivity to the transport models is left to future work.

\section{Conclusions and future work}
\label{sec:conclusions}

We have presented an accelerated Newton--Krylov solver for a $\theta$-implicit hybrid particle-in-cell formulation in which the electric field of the generalized Ohm's law and the electron pressure are the Newton unknowns and the particles are nonlinearly eliminated. Two components reduce the cost of the solve. A physics-based multigrid preconditioner extends the Hall-MHD preconditioner of Ref.~\cite{Chacon2025HMHD} to kinetic ions: it approximates the stiff field response, including the Hall, dissipative, electron-convection, and local ion responses, and relaxes it with a block-Jacobi smoother inside geometric multigrid. An analytical mass-matrix (AMM) approximation \cite{Chen2015Darwin} replaces the particle-based mass matrices in the Jacobian-vector product by a moment approximation computed from the grid charge density and magnetic field. The analytical mass matrices involve no particles and store only four reals per node, resulting in significant complexity and memory storage reductions. Because both enter only the Jacobian approximation and not the nonlinear residual, the converged solution and its conservation of energy and momentum are unchanged.

The solver reproduces parallel whistler and oblique kinetic Alfv\'en waves to better than $0.1\%$ in frequency, with second-order convergence in space and time, and the normal-mode spectrum of a plasma-filled conducting cylinder. On the GEM reconnection problem, it follows the fully implicit reference of Stanier et al.~\cite{Stanier2019}, and with the Newton tolerance tightened to $10^{-13}$ it conserves total energy to roundoff and total momentum to better than $10^{-9}$ in Alfv\'enic units. At $\Delta t\Omega_{ci}=0.05$, 
the combination of the preconditioner and the analytical mass matrices reduce the time to solution by 186-fold relative to unpreconditioned JFNK. Over the full reconnection interval, an explicit hybrid-PIC calculation at the implicit time step is faster, but its total-energy error is more than four orders of magnitude larger; reaching a comparable energy error requires a forty-times-smaller time step and nine times the wall-clock time. The robustness and applicability of the solver for highly non-linear problems is demonstrated by simulating the formation of a field-reversed configuration, which involves rapidly varying external fields, vacuum regions, and magnetic reconnection.

Future work will extend the algorithm to include anisotropic electron heat conduction, whose restrictive explicit time step~\cite{Sharma2011Diffusion} makes it a natural target for the implicit solver and its pressure preconditioner (e.g., using the scheme proposed in Ref.~\cite{chacon2025robust}).

\section*{CRediT authorship contribution statement}

\textbf{Prabhat Kumar:} Conceptualization, Methodology, Software, Validation, Writing -- original draft. \textbf{Luis Chac\'on:} Conceptualization, Methodology, Validation, Writing -- review \& editing. \textbf{William J. White:} Investigation, Formal analysis, Writing -- original draft. \textbf{Lujain Almarhabi:} Investigation, Formal analysis, Writing -- original draft. \textbf{Stephen E. Clark:} Conceptualization, Methodology, Software, Formal analysis, Writing -- review \& editing. 

\section*{Declaration of competing interest}

The authors declare that they have no known competing financial interests or personal relationships that could have appeared to influence the work reported in this paper.

\section*{Acknowledgments}

This research used resources of the National Energy Research Scientific Computing Center (NERSC), a Department of Energy Office of Science User Facility using NERSC award FES-ERCAP32666. This research used the open-source particle-in-cell code WarpX. Primary WarpX contributors are with LBNL, LLNL, CEA-LIDYL, SLAC, DESY, CERN, Helion Energy, TAE Technologies, and Realta Fusion. We acknowledge all WarpX contributors.

\section*{Data availability}

Data will be made available on request.

\section*{Declaration of generative AI and AI-assisted technologies in the manuscript preparation process}

During the preparation of this work the author(s) used Claude Fable 5.1 (Anthropic) to assist in implementing the algorithm and refining the manuscript. After using this tool/service, the author(s) reviewed and edited the content as needed and take(s) full responsibility for the content of the published article.

\appendix
\appendix
\section{Derivation of the hybrid-PIC preconditioner blocks}
\label{app:pcderivation}

This appendix supplies the algebra specific to the electric-field formulation and its kinetic-ion and pressure responses. The auxiliary-current split and the local block factorization follow Ref.~\cite{Chacon2025HMHD} (Sec.~4.1 and Appendix~C); the discrete energy derivation is in the appendix of Ref.~\cite{Kumar2026ThetaImplicitHybrid}.

\subsection{Field update and the ion--dissipation coupling}
\label{app:fieldchain}

Write $\tau=\theta\Delta t$, $\vect{e}=\dE$, and $p=\delta P_e$. Before the coefficients are frozen, the update of the field residual is
\begin{equation}
\begin{aligned}
 \delta\vect{F}_E={}&\vect{e}
 -\frac{(\delta\J-\delta\Ji)\times\B_0
              +\J_{e,0}\times\delta\B}{\rho_{c,0}}
 +\frac{\nabla_h p}{\rho_{c,0}}-\delta\vect{D}\\
 &+\frac{\delta\rhoc}{\rho_{c,0}^{\,2}}
       (\J_{e,0}\times\B_0-\nabla_h P_{e,0}).
\end{aligned}
\label{eq:appfullfield}
\end{equation}
The update of the dissipative field is
\begin{equation}
 \delta\vect{D}=\eta_0\delta\J-\eta_{h,0}\nabla_h^2\delta\J
 +\delta\eta\,\J_0-\delta\eta_h\nabla_h^2\J_0.
 \label{eq:appdissvariation}
\end{equation}
The preconditioner drops the density-update term in Eq.~\eqref{eq:appfullfield} and the coefficient updates in Eq.~\eqref{eq:appdissvariation}. Faraday's and Ampère's laws then give
\begin{equation}
 \delta\B=-\tau\nabla_h\times\vect{e},\qquad
 \delta\J=-\frac{\tau}{\mu_0}\Kop\vect{e},\qquad
 \delta\vect{D}=-\Dop\Kop\vect{e}.
 \label{eq:appresponsechain}
\end{equation}
The same-node approximation of the ion response acts on the push-field update, giving
\begin{equation}
 \delta\E^*=\vect{e}+\Dop\Kop\vect{e},\qquad
 \delta\Ji\simeq\boldsymbol{\Sigma}
                (\vect{e}+\Dop\Kop\vect{e}).
 \label{eq:apppushresponse}
\end{equation}
Using the definitions of $\Hop$, $\Smat$, and $\vect{w}$, the four field contributions in Eq.~\eqref{eq:appfullfield} collect as
\begin{equation}
\begin{aligned}
 \delta\vect{F}_E\simeq{}&\vect{e}
 +\Hop\Kop\vect{e}
 +\Smat(\vect{e}+\Dop\Kop\vect{e})
 +\Dop\Kop\vect{e}
 -\vect{w}\times(\nabla_h\times\vect{e})+G_Pp\\
 ={}&\Amat\vect{e}+(\Hop+\Dop+\Smat\Dop)\Kop\vect{e}
 -\vect{w}\times(\nabla_h\times\vect{e})+G_Pp.
\end{aligned}
\label{eq:appcollectedfield}
\end{equation}
The auxiliary variable $\vect{W}=\Kop\vect{e}=-\mu_0\delta\J/\tau$ is proportional to the plasma-current update, which is why it takes the boundary parities of the current.

\subsection{Local block and closed-form inverse}
\label{app:blockjacobi}

The local block $\mathsf{B}_g$ of Section~\ref{subsec:fieldblock} approximates the self-coupling of node $g$ in $\mathcal{A}$. Derivative couplings are replaced by the scales $\ell_1$ and $\ell_2$, $\Amat^{-1}$ by the identity, and $\boldsymbol{\Sigma}$ by its isotropic mean, so that the block has a closed-form inverse; the residual retains $\Amat^{-1}$ and the anisotropic $\boldsymbol{\Sigma}$. We give the Cartesian form; on the cylindrical mesh, the scales of the cylindrical operators replace $\ell_1$ and $\ell_2$. With the projectors $\mathsf{P}_{\parallel}=\bhat\bhat^T$ and $\mathsf{P}_{\perp}=\I-\mathsf{P}_{\parallel}$ and $\Cb\vect{v}=\vect{v}\times\bhat$, the local forms of $\Kop$ and of $\Hop+\Dop+\Smat\Dop$ are
\begin{equation}
 \mathsf{R}=\ell_2(\mathsf{P}_{\parallel}+\chi_w\mathsf{P}_{\perp}),
 \qquad U_{\rm loc}=a_U\I+c_{\rm eff}\Cb,
 \label{eq:appR}
\end{equation}
with
\begin{equation}
 a_U=a_\eta+a_h\ell_2,\qquad
 c_{\rm eff}=c_w+\overline{m}\,a_U.
 \label{eq:localcoeffs}
\end{equation}
The factor $\chi_w$ reduces the perpendicular part of the curl--curl scale, since the centered composition $\Kop$ peaks at $\ell_2/2$ on the Cartesian mesh. With the diagonal block
\begin{equation}
 \mathsf{T}=(1+|\vect{w}|\ell_1+a_h\ell_2^2)\I
 +(1-\chi_w)\ell_2a_U\mathsf{P}_{\perp},
 \label{eq:appT}
\end{equation}
the local system for the electric and scaled auxiliary updates $(\vect{e},\vect{h})$ is
\begin{equation}
 \mathsf{B}_g\begin{bmatrix}\vect{e}\\\vect{h}\end{bmatrix}
 =\begin{bmatrix}
 \mathsf{T}&s_W^{-1}U_{\rm loc}\\
 -s_W\mathsf{R}&\I
 \end{bmatrix}
 \begin{bmatrix}\vect{e}\\\vect{h}\end{bmatrix}
 =\begin{bmatrix}\vect{r}_E\\\vect{r}_W\end{bmatrix}.
 \label{eq:applocalblock}
\end{equation}
Eliminating its second row gives
\begin{equation}
 \vect{h}=\vect{r}_W+s_W\mathsf{R}\vect{e},\qquad
 (\mathsf{T}+U_{\rm loc}\mathsf{R})\vect{e}
 =\vect{r}_E-s_W^{-1}U_{\rm loc}\vect{r}_W.
 \label{eq:appschur}
\end{equation}
Since $\Cb\mathsf{P}_{\parallel}=0$ and $\Cb\mathsf{P}_{\perp}=\Cb$, the Schur complement is
\begin{equation}
\begin{aligned}
 \mathsf{T}+U_{\rm loc}\mathsf{R}
 &=[1+|\vect{w}|\ell_1+a_h\ell_2^2+\ell_2a_U]\I
   +\chi_w\ell_2c_{\rm eff}\Cb\\
 &=\alpha\I+\beta\Cb,
\end{aligned}
\label{eq:appschurcoeffs}
\end{equation}
with
\begin{equation}
 \alpha=1+|\vect{w}|\ell_1+\ell_2(a_\eta+2\ell_2a_h),\qquad
 \beta=\chi_w\ell_2c_{\rm eff}.
 \label{eq:alpha}
\end{equation}
Three terms in $\mathsf{T}$ set the effectiveness of the block as a smoother. The term $|\vect{w}|\ell_1$ gives the convection a positive local scale. The term $a_h\ell_2^2$ makes the hyper-resistive part of $\alpha$ use $2\ell_2$, the largest eigenvalue of the compact Laplacian. The term $(1-\chi_w)\ell_2a_U\mathsf{P}_{\perp}$ restores the full scale $\ell_2$ to the dissipative part, so that $\chi_w$ reduces only the rotation coefficient $\beta$.

For a unit $\bhat$, the identity $\Cb^2=-\mathsf{P}_{\perp}$ separates the inverse into parallel and perpendicular parts:
\begin{equation}
 (\alpha\I+\beta\Cb)^{-1}
 =\frac{\mathsf{P}_{\parallel}}{\alpha}
 +\frac{\alpha\mathsf{P}_{\perp}-\beta\Cb}{\alpha^2+\beta^2}.
 \label{eq:appprojectorinverse}
\end{equation}
Expanding the projectors gives
\begin{equation}
 (\alpha\I+\beta\Cb)^{-1}\vect{q}
 =\frac{\alpha\vect{q}-\beta(\vect{q}\times\bhat)
 +(\beta^2/\alpha)(\vect{q}\cdot\bhat)\bhat}{\alpha^2+\beta^2},
 \label{eq:localinverse}
\end{equation}
and Eq.~\eqref{eq:Ainv} is the same identity with $\alpha=1$ and $\beta=\overline{m}$. At a magnetic null, we set $\bhat=\vect{0}$; then $\Cb=0$, $\mathsf{R}=\chi_w\ell_2\I$, and the inverse is $\I/\alpha$. The Schur reduction and rotation inverse are the algebra of Ref.~\cite{Chacon2025HMHD} (Appendix~C, Eqs.~(C.3)--(C.4)), extended here by the local kinetic-ion response and the stabilization above.

\subsection{Pressure update and transport--conduction stencil}
\label{app:pressure}

Let $\mathcal{T}_h(\vect{u}_c,P_e)$ denote the conservative limited-flux divergence used for pressure transport, and again set $\tau=\theta\Delta t$. The normalized pressure residual is
\begin{equation}
\begin{aligned}
 F_P={}&P_e-P_e^n+\tau\gamma\mathcal{T}_h(\vect{u}_c,P_e)\\
 &-\tau(\gamma-1)\left[\Je\cdot\E^*+Q_e
 +\nabla_h\cdot\left(\kappa_e\nabla_h\frac{P_e}{n_e}\right)\right].
\end{aligned}
\label{eq:apppressureresidual}
\end{equation}
Writing $n_0=n_{e,0}$ and $P_0=P_{e,0}$, its update is
\begin{equation}
\begin{aligned}
 \delta F_P={}&p+\tau\gamma\,\delta\mathcal{T}_h\\
 &-\tau(\gamma-1)\bigg[\delta\Je\cdot\E_0^*
 +\J_{e,0}\cdot(\vect{e}-\delta\vect{D})+\delta Q_e\\
 &\qquad+\nabla_h\cdot\bigg\{\kappa_{e,0}\nabla_h
       \left(\frac{p}{n_0}-\frac{P_0\delta n_e}{n_0^2}\right)
       +\delta\kappa_e\nabla_h\frac{P_0}{n_0}\bigg\}\bigg].
\end{aligned}
\label{eq:apppressurevariation}
\end{equation}
The velocity update within $\delta\mathcal{T}_h$, the work and heating updates, and the density response contribute to $C_{PE}$ in Eq.~\eqref{eq:coupledblocks}. At fixed electric field, these contributions vanish for the model considered here. A temperature-dependent conductivity can still vary through $p/n_0$; freezing this remaining response leaves
\begin{equation}
 J_{PP}^{\rm fr}p
 =p+\tau\gamma\mathcal{T}'_{h,P}p
 -\tau(\gamma-1)\nabla_h\cdot
       \left[\kappa_{e,0}\nabla_h\left(\frac{p}{n_0}\right)\right],
 \label{eq:apppressurefrozen}
\end{equation}
where $\mathcal{T}'_{h,P}$ is the pressure derivative of the limited conservative flux at fixed velocity. The pressure preconditioner replaces this operator by the advection and diffusion parts in Eq.~\eqref{eq:pressureop}. In continuum notation the omitted transport contribution includes $\gamma(\nabla\cdot\vect{u}_c)p$. Similarly, replacing the heat-flux operator by $(\kappa_0/n_0)\nabla_h^2p$ omits coefficient-gradient contributions and uses the reference conductivity rather than its full spatial and thermodynamic dependence.

For an interior Cartesian node, define
\begin{equation}
 a_{d,g}=\tau\gamma\frac{|u_{c,d,g}|}{\Delta x_d},\qquad
 c_{\kappa,g}=\tau(\gamma-1)\frac{\kappa_0}{n_{0,g}}.
 \label{eq:apppressureweights}
\end{equation}
With $g_d^{\rm up}=g-\operatorname{sgn}(u_{c,d,g})\hat d$, the stencil is
\begin{equation}
\begin{aligned}
 (L_Pp)_g={}&\left(1+\sum_d a_{d,g}+c_{\kappa,g}\sum_d\frac{2}{\Delta x_d^2}\right)p_g
 -\sum_d a_{d,g}p_{g_d^{\rm up}}\\
 &-\sum_d\frac{c_{\kappa,g}}{\Delta x_d^2}
             (p_{g+\hat d}+p_{g-\hat d}).
\end{aligned}
\label{eq:apppressurestencil}
\end{equation}
The coefficient of $p_g$ gives Eq.~\eqref{eq:Dp}, and the remaining terms supply the Jacobi sweeps.

\bibliographystyle{elsarticle-num}
\bibliography{refs}
\end{document}